\input epsf
%
%
%
\def\unredoffs{} 

%
%
%
%
\newbox\leftpage \newdimen\fullhsize \newdimen\hstitle \newdimen\hsbody
\tolerance=1000\hfuzz=2pt
\catcode`\@=11 
%
\magnification=1200\unredoffs\baselineskip=16pt plus 2pt minus 1pt
\hsbody=\hsize \hstitle=\hsize 
%
%
%
\newcount\yearltd\yearltd=\year\advance\yearltd by -1900

%
%

\def\draftmode{\message{ DRAFTMODE }\def\draftdate{{\rm preliminary draft:
\number\month/\number\day/\number\yearltd\ \ \hourmin}}%
\headline={\hfil\draftdate}\writelabels\baselineskip=20pt plus 2pt minus 2pt
 {\count255=\time\divide\count255 by 60 \xdef\hourmin{\number\count255}
  \multiply\count255 by-60\advance\count255 by\time
  \xdef\hourmin{\hourmin:\ifnum\count255<10 0\fi\the\count255}}}
\def\nolabels{\def\wrlabeL##1{}\def\eqlabeL##1{}\def\reflabeL##1{}}
\def\writelabels{\def\wrlabeL##1{\leavevmode\vadjust{\rlap{\smash%
{\line{{\escapechar=` \hfill\rlap{\sevenrm\hskip.03in\string##1}}}}}}}%
\def\eqlabeL##1{{\escapechar-1\rlap{\sevenrm\hskip.05in\string##1}}}%
\def\reflabeL##1{\noexpand\llap{\noexpand\sevenrm\string\string\string##1}}}
\nolabels
%
\global\newcount\secno \global\secno=0
\global\newcount\meqno \global\meqno=1
\def\newsec#1{\global\advance\secno by1\message{(\the\secno. #1)}
\global\subsecno=0\eqnres@t\noindent{\bf\the\secno. #1}
\writetoca{{\secsym} {#1}}\par\nobreak\medskip\nobreak}
\def\eqnres@t{\xdef\secsym{\the\secno.}\global\meqno=1\bigbreak\bigskip}
\def\sequentialequations{\def\eqnres@t{\bigbreak}}\xdef\secsym{}
\global\newcount\subsecno \global\subsecno=0
\def\subsec#1{\global\advance\subsecno by1\message{(\secsym\the\subsecno. #1)}
\ifnum\lastpenalty>9000\else\bigbreak\fi
\noindent{\it\secsym\the\subsecno. #1}\writetoca{\string\quad
{\secsym\the\subsecno.} {#1}}\par\nobreak\medskip\nobreak}
\def\appendix#1#2{\global\meqno=1\global\subsecno=0\xdef\secsym{\hbox{#1.}}
\bigbreak\bigskip\noindent{\bf Appendix #1. #2}\message{(#1. #2)}
\writetoca{Appendix {#1.} {#2}}\par\nobreak\medskip\nobreak}
%
%
\def\eqnn#1{\xdef #1{(\secsym\the\meqno)}\writedef{#1\leftbracket#1}%
\global\advance\meqno by1\wrlabeL#1}
\def\eqna#1{\xdef #1##1{\hbox{$(\secsym\the\meqno##1)$}}
\writedef{#1\numbersign1\leftbracket#1{\numbersign1}}%
\global\advance\meqno by1\wrlabeL{#1$\{\}$}}
\def\eqn#1#2{\xdef #1{(\secsym\the\meqno)}\writedef{#1\leftbracket#1}%
\global\advance\meqno by1$$#2\eqno#1\eqlabeL#1$$}
%
\newskip\footskip\footskip14pt plus 1pt minus 1pt 
\def\footnotefont{\ninepoint}\def\f@t#1{\footnotefont #1\@foot}
\def\f@@t{\baselineskip\footskip\bgroup\footnotefont\aftergroup\@foot\let\next}
\setbox\strutbox=\hbox{\vrule height9.5pt depth4.5pt width0pt}
\global\newcount\ftno \global\ftno=0
\def\foot{\global\advance\ftno by1\footnote{$^{\the\ftno}$}}
%
\newwrite\ftfile
\def\footend{\def\foot{\global\advance\ftno by1\chardef\wfile=\ftfile
$^{\the\ftno}$\ifnum\ftno=1\immediate\openout\ftfile=foots.tmp\fi%
\immediate\write\ftfile{\noexpand\smallskip%
\noexpand\item{f\the\ftno:\ }\pctsign}\findarg}%
\def\footatend{\vfill\eject\immediate\closeout\ftfile{\parindent=20pt
\centerline{\bf Footnotes}\nobreak\bigskip\input foots.tmp }}}
\def\footatend{}
%
%
\global\newcount\refno \global\refno=1
\newwrite\rfile
\def\ref{[\the\refno]\nref}
\def\nref#1{\xdef#1{[\the\refno]}\writedef{#1\leftbracket#1}%
\ifnum\refno=1\immediate\openout\rfile=refs.tmp\fi
\global\advance\refno by1\chardef\wfile=\rfile\immediate
\write\rfile{\noexpand\item{#1\ }\reflabeL{#1\hskip.31in}\pctsign}\findarg}
\def\findarg#1#{\begingroup\obeylines\newlinechar=`\^^M\pass@rg}
{\obeylines\gdef\pass@rg#1{\writ@line\relax #1^^M\hbox{}^^M}%
\gdef\writ@line#1^^M{\expandafter\toks0\expandafter{\striprel@x #1}%
\edef\next{\the\toks0}\ifx\next\em@rk\let\next=\endgroup\else\ifx\next\empty%
\else\immediate\write\wfile{\the\toks0}\fi\let\next=\writ@line\fi\next\relax}}
\def\striprel@x#1{} \def\em@rk{\hbox{}}
\def\lref{\begingroup\obeylines\lr@f}
\def\lr@f#1#2{\gdef#1{\ref#1{#2}}\endgroup\unskip}
\def\semi{;\hfil\break}
\def\addref#1{\immediate\write\rfile{\noexpand\item{}#1}} 
\def\startrefs#1{\immediate\openout\rfile=refs.tmp\refno=#1}
\def\xref{\expandafter\xr@f}\def\xr@f[#1]{#1}
\def\refs#1{\count255=1[\r@fs #1{\hbox{}}]}
\def\r@fs#1{\ifx\und@fined#1\message{reflabel \string#1 is undefined.}%
\nref#1{need to supply reference \string#1.}\fi%
\vphantom{\hphantom{#1}}\edef\next{#1}\ifx\next\em@rk\def\next{}%
\else\ifx\next#1\ifodd\count255\relax\xref#1\count255=0\fi%
\else#1\count255=1\fi\let\next=\r@fs\fi\next}
%

%
\newwrite\ffile\global\newcount\figno \global\figno=1
\def\fig{fig.~\the\figno\nfig}
\def\nfig#1{\xdef#1{fig.~\the\figno}%
\writedef{#1\leftbracket fig.\noexpand~\the\figno}%
\ifnum\figno=1\immediate\openout\ffile=figs.tmp\fi\chardef\wfile=\ffile%
\immediate\write\ffile{\noexpand\medskip\noexpand\item{Fig.\ \the\figno. }
\reflabeL{#1\hskip.55in}\pctsign}\global\advance\figno by1\findarg}
\def\xfig{\expandafter\xf@g}\def\xf@g fig.\penalty\@M\ {}
\def\figs#1{figs.~\f@gs #1{\hbox{}}}
\def\f@gs#1{\edef\next{#1}\ifx\next\em@rk\def\next{}\else
\ifx\next#1\xfig #1\else#1\fi\let\next=\f@gs\fi\next}
\newwrite\lfile
{\escapechar-1\xdef\pctsign{\string\%}\xdef\leftbracket{\string\{}
\xdef\rightbracket{\string\}}\xdef\numbersign{\string\#}}

\def\writestop{\def\writestoppt{\immediate\write\lfile{\string\pageno%
\the\pageno\string\startrefs\leftbracket\the\refno\rightbracket%
\string\def\string\secsym\leftbracket\secsym\rightbracket%
\string\secno\the\secno\string\meqno\the\meqno}\immediate\closeout\lfile}}
\def\writestoppt{}\def\writedef#1{}
\def\seclab#1{\xdef #1{\the\secno}\writedef{#1\leftbracket#1}\wrlabeL{#1=#1}}
\def\subseclab#1{\xdef #1{\secsym\the\subsecno}%
\writedef{#1\leftbracket#1}\wrlabeL{#1=#1}}
\newwrite\tfile \def\writetoca#1{}
\def\leaderfill{\leaders\hbox to 1em{\hss.\hss}\hfill}
\def\writetoc{\immediate\openout\tfile=toc.tmp
   \def\writetoca##1{{\edef\next{\write\tfile{\noindent ##1
   \string\leaderfill {\noexpand\number\pageno} \par}}\next}}}

%
\catcode`\@=12 
%
\edef\tfontsize{\ifx\answ\bigans scaled\magstep3\else scaled\magstep4\fi}
\font\titlerm=cmr10 \tfontsize \font\titlerms=cmr7 \tfontsize
\font\titlermss=cmr5 \tfontsize \font\titlei=cmmi10 \tfontsize
\font\titleis=cmmi7 \tfontsize \font\titleiss=cmmi5 \tfontsize
\font\titlesy=cmsy10 \tfontsize \font\titlesys=cmsy7 \tfontsize
\font\titlesyss=cmsy5 \tfontsize \font\titleit=cmti10 \tfontsize
\skewchar\titlei='177 \skewchar\titleis='177 \skewchar\titleiss='177
\skewchar\titlesy='60 \skewchar\titlesys='60 \skewchar\titlesyss='60
\def\titlefont{\def\rm{\fam0\titlerm}
\textfont0=\titlerm \scriptfont0=\titlerms \scriptscriptfont0=\titlermss
\textfont1=\titlei \scriptfont1=\titleis \scriptscriptfont1=\titleiss
\textfont2=\titlesy \scriptfont2=\titlesys \scriptscriptfont2=\titlesyss
\textfont\itfam=\titleit \def\it{\fam\itfam\titleit}\rm}
 \ifx\answ\bigans\else scaled\magstep1\fi
\ifx\answ\bigans\else

 \font\absi=cmmi10 scaled\magstep1
\font\absis=cmmi7 scaled\magstep1 \font\absiss=cmmi5 scaled\magstep1
\font\abssy=cmsy10 scaled\magstep1 \font\abssys=cmsy7 scaled\magstep1
\font\abssyss=cmsy5 scaled\magstep1 
\skewchar\absi='177 \skewchar\absis='177 \skewchar\absiss='177
\skewchar\abssy='60 \skewchar\abssys='60 \skewchar\abssyss='60
\fi
\font\ninerm=cmr9 \font\sixrm=cmr6 \font\ninei=cmmi9 \font\sixi=cmmi6
\font\ninesy=cmsy9 \font\sixsy=cmsy6 \font\ninebf=cmbx9
\font\nineit=cmti9 \font\ninesl=cmsl9 \skewchar\ninei='177
\skewchar\sixi='177 \skewchar\ninesy='60 \skewchar\sixsy='60
\def\ninepoint{\def\rm{\fam0\ninerm}
\textfont0=\ninerm \scriptfont0=\sixrm \scriptscriptfont0=\fiverm
\textfont1=\ninei \scriptfont1=\sixi \scriptscriptfont1=\fivei
\textfont2=\ninesy \scriptfont2=\sixsy \scriptscriptfont2=\fivesy
\textfont\itfam=\ninei \def\it{\fam\itfam\nineit}\def\sl{\fam\slfam\ninesl}%
\textfont\bffam=\ninebf \def\bf{\fam\bffam\ninebf}\rm}
%
%
\def\noblackbox{\overfullrule=0pt}
\hyphenation{anom-aly anom-alies coun-ter-term coun-ter-terms}
\def\inv{^{\raise.15ex\hbox{${\scriptscriptstyle -}$}\kern-.05em 1}}

\def\Dsl{\,\raise.15ex\hbox{/}\mkern-13.5mu D} 
\def\dsl{\raise.15ex\hbox{/}\kern-.57em\partial}

\def\lspace{\ifx\answ\bigans{}\else\qquad\fi}
\def\lbspace{\ifx\answ\bigans{}\else\hskip-.2in\fi} 
\def\boxeqn#1{\vcenter{\vbox{\hrule\hbox{\vrule\kern3pt\vbox{\kern3pt
        \hbox{${\displaystyle #1}$}\kern3pt}\kern3pt\vrule}\hrule}}}
\def\mbox#1#2{\vcenter{\hrule \hbox{\vrule height#2in
                \kern#1in \vrule} \hrule}}  
%
   
 \def\CH{{\cal H}}

\def\darr#1{\raise1.5ex\hbox{$\leftrightarrow$}\mkern-16.5mu #1}
\def\chipp{\kappa}

\def\half{{\textstyle{1\over2}}} 
\def\roughly#1{\raise.3ex\hbox{$#1$\kern-.75em\lower1ex\hbox{$\sim$}}}
\hyphenation{Mar-ti-nel-li}

\def\Re{\,\hbox{Re}\,}

\def\1{\;1\!\!\!\! 1\;}

\def\etal{{\it et al.}}

\def\toinf#1{\mathrel{\mathop{\sim}\limits_{\scriptscriptstyle
{#1\rightarrow\infty }}}}
\def\tozero#1{\mathrel{\mathop{\sim}\limits_{\scriptscriptstyle
{#1\rightarrow0 }}}}
\def\frac#1#2{{{#1}\over {#2}}}
\def\half{\hbox{${1\over 2}$}}
\def\quarter{\hbox{${1\over 4}$}}
\def\smallfrac#1#2{\hbox{${{#1}\over {#2}}$}}

\def\MS{\hbox{$\overline{\rm MS}$}}

\catcode`@=11 
\def\slash#1{\mathord{\mathpalette\c@ncel#1}}
 \def\c@ncel#1#2{\ooalign{$\hfil#1\mkern1mu/\hfil$\crcr$#1#2$}}
\def\lsim{\mathrel{\mathpalette\@versim<}}
\def\gsim{\mathrel{\mathpalette\@versim>}}
 \def\@versim#1#2{\lower0.2ex\vbox{\baselineskip\z@skip\lineskip\z@skip
       \lineskiplimit\z@\ialign{$\m@th#1\hfil##$\crcr#2\crcr\sim\crcr}}}
\catcode`@=12 

\def\PR{{\it Phys.~Rev.~}}

\def\NP{{\it Nucl.~Phys.~}}

\def\PL{{\it Phys.~Lett.~}}

\def\SJNP{{\it Sov.~Jour.~Nucl.~Phys.~}}
\def\SPJETP{{\it Sov.~Phys.~J.E.T.P.~}}

\def\JHEP{{\it Jour.~High~Energy~Phys.~}}
\def\vol#1{{\bf #1}}\def\vyp#1#2#3{\vol{#1} (#2) #3}

\def\as{\alpha_s}

\def\Ai{\hbox{Ai}}
\def\ash{\widehat\alpha_s}

\noblackbox
\pageno=0\nopagenumbers\tolerance=10000\hfuzz=5pt
\baselineskip 12pt
\line{\hfill {\tt hep-ph/0109178}}
\line{\hfill CERN-TH/2001-206}
\line{\hfill Edinburgh 2001-10}
\line{\hfill RM3-TH/01-08}
\vskip 12pt
\centerline{\titlefont Factorization and  Resummation of }\vskip 10pt
\centerline{\titlefont Small $x$  Scaling Violations with Running Coupling}
\vskip 18pt\centerline{Guido~Altarelli,$^{(a)}$ 
Richard D.~Ball$^{(a,b)}$ and Stefano Forte$^{(c),}$\footnote{$^\dagger$}{On leave from INFN,
Sezione di Torino, Italy}}
\vskip 12pt
\centerline{\it ${}^{(a)}$Theory Division, CERN,}
\centerline{\it CH-1211 Gen\`eve 23, Switzerland.}
\vskip 6pt
\centerline{\it ${}^{(b)}$Department of Physics and Astronomy}
\centerline{\it University of Edinburgh, Edinburgh EH9 3JZ, Scotland}
\vskip 6pt
\centerline {\it ${}^{(c)}$INFN, Sezione di Roma III}
\centerline {\it Via della Vasca Navale 84, I-00146 Rome, Italy}
\vskip 50pt
\centerline{\bf Abstract}
{\narrower\baselineskip 10pt
\medskip\noindent
We discuss the inclusion of running  coupling effects
in perturbative small $x$ evolution equations. We
show that a running coupling BFKL--like $x$--evolution
equation is fully compatible, up to higher twist corrections, with 
the standard factorized perturbative evolution of parton
distributions. We then use this
result, combined with the well--known Airy asymptotics, to prove that
the oscillations which are present in the running--coupling BFKL
solution do not affect the associated splitting functions, which
instead remain smooth in the small $x$ limit.
This allows us to give a prescription to include 
running--coupling corrections
in the small--$x$ resummation of scaling violations. We show that these
corrections are small in the HERA region.}
\vfill
\line{CERN-TH/2001-206\hfill }
\line{September 2001\hfill}
\eject \footline={\hss\tenrm\folio\hss}

\lref\glap{ 
V.N.~Gribov and L.N.~Lipatov, 
\SJNP\vyp{15}{1972}{438}\semi  
L.N.~Lipatov, \SJNP\vyp{20}{1975}{95}\semi    
G.~Altarelli and G.~Parisi, 
\NP\vyp{B126}{1977}{298}\semi  
see also
Y.L.~Dokshitzer, 
{\it Sov.~Phys.~JETP~}\vyp{46}{1977}{691}.} 
\lref\nlo{G.~Curci, W.~Furma\'nski and R.~Petronzio, 
\NP\vyp{B175}{1980}{27}\semi 
E.G.~Floratos, C.~Kounnas and R.~Lacaze, 
\NP\vyp{B192}{1981}{417}.} 
\lref\nnlo{S.A.~Larin, T.~van~Ritbergen, J.A.M.~Vermaseren,
\NP\vyp{B427}{1994}{41}\semi  
S.A.~Larin \etal, \NP\vyp{B492}{1997}{338}.} 

\lref\bfkl{L.N.~Lipatov, 
\SJNP\vyp{23}{1976}{338}\semi 
 V.S.~Fadin, E.A.~Kuraev and L.N.~Lipatov, 
\PL\vyp{60B}{1975}{50}; 
 {\it Sov. Phys. JETP~}\vyp{44}{1976}{443}; 
\vyp{45}{1977}{199}\semi 
 Y.Y.~Balitski and L.N.Lipatov, 
\SJNP\vyp{28}{1978}{822}.} 
\lref\CH{ 
S.~Catani and F.~Hautmann, 
\PL\vyp{B315}{1993}{157}; 
\NP\vyp{B427}{1994}{475}.} 
\lref\fl{V.S.~Fadin and L.N.~Lipatov, 
\PL\vyp{B429}{1998}{127}\semi  
V.S.~Fadin et al, \PL\vyp{B359}{1995}{181}; 
\PL\vyp{B387}{1996}{593}; 
\NP\vyp{B406}{1993}{259}; 
\PR\vyp{D50}{1994}{5893}; 
\PL\vyp{B389}{1996}{737};  
\NP\vyp{B477}{1996}{767};  
\PL\vyp{B415}{1997}{97};  
\PL\vyp{B422}{1998}{287}.} 
\lref\cc{G.~Camici and M.~Ciafaloni, 
\PL\vyp{B412}{1997}{396}; 
\PL\vyp{B430}{1998}{349}.} 
\lref\dd{V.~del~Duca, \PR\vyp{D54}{1996}{989};
\PR\vyp{D54}{1996}{4474}\semi 
V.~del~Duca and C.R.~Schmidt, 
\PR\vyp{D57}{1998}{4069}\semi 
Z.~Bern, V.~del~Duca and C.R.~Schmidt, 
\PL\vyp{B445}{1998}{168}.}
\lref\ross{
D.~A.~Ross,
Phys.\ Lett.\ B {\bf 431}, 161 (1998) 
}
\lref\jar{T.~Jaroszewicz, 
\PL\vyp{B116}{1982}{291}.}
\lref\ktfac{S.~Catani, F.~Fiorani and G.~Marchesini, 
\PL\vyp{B336}{1990}{18}\semi 
S.~Catani et al., 
\NP\vyp{B361}{1991}{645}.}
\lref\summ{R.~D.~Ball and S.~Forte, 
\PL\vyp{B351}{1995}{313}\semi  
R.K.~Ellis, F.~Hautmann and B.R.~Webber, 
\PL\vyp{B348}{1995}{582}.}
\lref\afp{R.~D.~Ball and S.~Forte, 
\PL\vyp{B405}{1997}{317}.}
\lref\DGPTWZ{A.~De~R\'ujula {\it et al.}, 
\PR\vyp{D10}{1974}{1649}.}
\lref\das{R.~D.~Ball and S.~Forte, 
\PL\vyp{B335}{1994}{77}; 
\vyp{B336}{1994}{77}\semi 
{\it Acta~Phys.~Polon.~}\vyp{B26}{1995}{2097}.}
\lref\kis{ 
See {\it e.g.}  R.~K.~Ellis, W.~J.~Stirling and B.~R.~Webber, 
``QCD and Collider Physics'' (C.U.P., Cambridge 1996).}
\lref\hone{H1 Collab., {\tt hep-ex/0012053}.}
\lref\zeus{ZEUS Collab., {\tt hep-ex/0105090}.} 
\lref\mom{R.~D.~Ball and S.~Forte, {\it Phys. Lett.} {\bf
B359}, 362 (1995).}
\lref\bfklfits{R.~D.~Ball and S.~Forte, 
{\tt hep-ph/9607291}\semi 
I.~Bojak and M.~Ernst, \PL\vyp{B397}{1997}{296};
\NP\vyp{B508}{1997}{731}\semi
J.~Bl\"umlein  and A.~Vogt, 
\PR\vyp{D58}{1998}{014020}.}
\lref\flph{R.~D.~Ball  and S.~Forte, 
{\tt hep-ph/9805315}\semi 
J. Bl\"umlein et al., 
{\tt hep-ph/9806368}.}
\lref\salam{G.~Salam, \JHEP\vyp{9807}{1998}{19}.}
\lref\sxap{R.~D.~Ball and S.~Forte, 
\PL\vyp{B465}{1999}{271}.}
\lref\sxres{G. Altarelli, R.~D. Ball and S. Forte, 
\NP{\bf B575}, 313 (2000);  
see also {\tt hep-ph/0001157}; 
}
\lref\sxphen{G. Altarelli, R.~D.~Ball and S. Forte,  
\NP\vyp{B599}{2001}{383};  
see also {\tt hep-ph/0104246}.}  
\lref\ciaf{M.~Ciafaloni and D.~Colferai, 
\PL\vyp{B452}{1999}{372}\semi 
M.~Ciafaloni, G.~Salam and  D.~Colferai, 
{\tt hep-ph/9905566}.}  
\lref\Liprun{L.N.~Lipatov, 
\SPJETP\vyp{63}{1986}{5}.}
\lref\ColKwie{ 
J.~C.~Collins and J.~Kwiecinski, \NP\vyp{B316}{1989}{307}.}
\lref\CiaMue{
M.~Ciafaloni, M.~Taiuti and A.~H.~Mueller,
{\tt hep-ph/0107009}.
}
\lref\ciafac{
M.~Ciafaloni, D.~Colferai and G.~P.~Salam,
JHEP {\bf 0007}  (2000) 054
}
\lref\ciafrun{G.~Camici and M.~Ciafaloni, 
\NP\vyp{B496}{1997}{305}.}
\lref\Haak{L.~P.~A.~Haakman, O.~V.~Kancheli and 
J.~H.~Koch \NP\vyp{B518}{1998}{275}.} 
\lref\KovMue{Y.V.~Kovchegov and A.H.~Mueller \PL\vyp{B439}{1998}{428}.} 
\lref\Bartels{N. Armesto, J. Bartels and M.~A.~Braun, 
\PL\vyp{B442}{1998}{459}.} 
\lref\Thorne{R.~S.~Thorne, 
\PL\vyp{B474}{2000}{372}.} 

\newsec{Introduction} 
\noindent 

The theory of scaling violations of deep inelastic structure 
functions at small $x$ has recently attracted considerable
interest, prompted by the extensive results obtained by 
experiments at HERA~\refs{\hone,\zeus}. New effects beyond the
low--order perturbative approximation~\refs{\glap\nlo{--}\nnlo} to
anomalous dimensions 
or splitting functions should become
important at small $x$. The BFKL approach~\refs{\bfkl\CH\fl\cc{--}\dd} 
provides in principle a tool for the 
determination of the small $x$
improvements of the anomalous dimensions~\refs{\jar,\ktfac}. 
However, the data are  
in good agreement with a standard
next--to--leading order perturbative treatment of 
scaling violations~\refs{\DGPTWZ,\das,\hone,\zeus}, while naive 
attempts~\summ\ to incorporate small $x$ logarithms fail 
completely~\refs{\bfklfits,\flph}. Important 
progress has been made recently in
understanding the empirical softness of the BFKL 
resummation~\refs{\salam\sxres\sxphen{--}\ciaf}. Specifically, it is now
understood that in order to avoid instabilities~\refs{\ross,\sxap},
the inclusion of small $x$ contributions must be
combined with a resummation of the collinear
singularities of standard perturbative evolution.

An important aspect of the problem has to do with the implementation 
of running coupling effects. Indeed, it is a priori not obvious that
the `duality' relations~\refs{\jar,\afp,\sxres} which connect BFKL evolution
in $\xi\equiv\ln\smallfrac{1}{x}$ and renormalization--group
evolution in $t\equiv\ln\smallfrac{Q^2}{\mu^2}$ remain
valid when the running of the coupling is taken into account.
These duality relations play a crucial role~\refs{\sxres,\sxphen}
in the desired simultaneous resummation of
collinear singularities and small $x$ logs, and 
imply that  in small--$x$ evolution factorization is preserved,
namely, anomalous dimensions remain independent of the boundary 
conditions on the parton density.

In our previous work~\refs{\sxap,\sxres,\sxphen}\ 
we have included running coupling effects perturbatively up to
NLLx. We have shown that duality and factorization continue
to hold, 
provided the anomalous dimension  is supplemented by a term
induced by the running of the coupling~\refs{\ciafrun,\sxap}.
This term however corresponds to a singular
contribution to the BFKL kernel, or equivalently, a perturbatively
unstable contribution to the anomalous dimension. In \MS\
factorization 
this singular contribution is shifted to the 
quark sector~\CH. Even though its effects are in practice rather small
in the HERA region~\sxphen, this singularity 
signals a failure of the NLLx treatment
of the running of the coupling in the asymptotic small $x$ limit.

In this paper, we address in detail the problems related to the
inclusion of running coupling effects in small $x$
evolution equations.
After reviewing the way duality is affected when the
running of the coupling is included perturbatively, we
consider  the modified running--coupling BFKL $x$--evolution equation
and its solution. We discuss the properties of the corresponding
anomalous dimension and prove factorization and duality to all
perturbative orders, up to higher twist corrections.~\foot{Arguments 
for factorization have already been presented elsewhere~\refs{\ciaf,\ciafac}, 
but they are partly based on model--dependent assumptions.}
We then study in detail the running
coupling contribution to anomalous dimensions
in the asymptotic small--$x$ limit, which may be determined exactly 
in terms of Airy functions, 
by taking a quadratic approximation of the BFKL kernel
near its minimum~\Liprun.
In particular, we prove that  the unphysical oscillatory
behaviour which has 
been shown~\refs{\ciaf,\Liprun\ColKwie\KovMue\Bartels\Thorne{--}\CiaMue} 
to characterize solutions to the
running--coupling BFKL equations at small $x$  
does not affect the small $x$ asymptotics
of splitting functions. Therefore, the perturbative approach
to scaling violations need not break down in the small $x$ limit.
 We then show  that this quadratic approximation is sufficient to
determine the leading 
$x\rightarrow 0$ asymptotics, since only the behaviour of the kernel near 
the minimum is relevant in this limit. 
Finally, we discuss how this asymptotic behaviour
can be systematically matched to the small $x$ expansion 
of anomalous dimensions, and
thus used to resum small $x$ running coupling terms. We show that the
resummed result is free of unphysical singularities, perturbatively
stable, and smooth in the small $x$ limit, and in the HERA region it
only leads to a small
correction to our previous phenomenological results.

\newsec{Perturbative Duality}

We start by reviewing the way duality of perturbative evolution is
proven, first at fixed coupling and then perturbatively   in the
running coupling case. We then show how the perturbative expansion in
the running coupling case is beset by unresummed singular contributions.
The behaviour of structure functions at small
$x$ is dominated by the large eigenvalue of evolution of the 
singlet parton component.  
Thus we consider the singlet parton density
\eqn\Gdef{ 
G(\xi,t)=x[g(x,Q^2)+k_q\otimes q(x,Q^2)],
} 
where $\xi=\log{1/x}$, $t=\log{Q^2/\mu^2}$, 
$g(x,Q^2)$ and $q(x,Q^2)$ are the gluon and singlet quark parton 
densities, respectively, and $k_q$ is such
that, for each moment
\eqn\Nmom{ 
G(N,t)=\int^{\infty}_{0}\! d\xi\, e^{-N\xi} G(\xi,t),
} 
the associated anomalous dimension $\gamma(\as(t),N)$ corresponds 
to the largest eigenvalue in the
singlet sector. The generalization to the full two--by--two matrix of
anomalous dimensions is discussed in detail in ref.~\sxphen. 

At large $t$ and fixed $\xi$ the evolution equation 
in $N$-moment space is then
\eqn\tevol{
\frac{d}{dt}G(N,t)=\gamma(\as(t),N) G(N,t),
} 
where $\as(t)$ is the running coupling. The anomalous dimension 
is completely known  
at one-- and two--loop level:
\eqn\gamexp{
\gamma(\as,N)=\as\gamma_0(N)+\as^2\gamma_1(N)+
\dots\>.
}
The corresponding splitting function is related by a  Mellin 
transform to $\gamma(\as,N)$:
\eqn\psasy{ 
\gamma(\as,N) =\int^1_0\!dx\,x^N\!P(\as,x)
} 
Small $x$ for the splitting function corresponds to small $N$ for 
the anomalous dimension: more precisely
$P\sim 1/x(\log(1/x))^n$ corresponds to $\gamma\sim n!/N^{n+1}$. 
Even assuming that a
leading twist description of scaling violations is still valid in 
some range of small $x$, as soon as
$x$ is small enough that 
$\as \xi\sim 1$, with $\xi=\log{1/x}$, all terms of order
$(\as/x) (\as \xi)^n$ (LLx) and $\as(\as/x) 
(\as \xi)^n$ (NLLx) 
which are present in the splitting
functions must be considered in order to achieve an accuracy up 
to terms of order
$\as^2(\as/x) (\as \xi)^n$ (NNLLx).  

As is well known, these terms can be derived from the knowledge of 
the kernel 
\eqn\chiexp{
\chi(\as,M)=\as\chi_0(M)+\as^2\chi_1(M)+\dots\>.
}
of the BFKL $\xi$--evolution equation 
\eqn\xevol{
\frac{d}{d\xi}G(\xi,M)=\chi(\as,M) G(\xi,M),
} 
which is satisfied at large $\xi$ by the inverse Mellin 
transform of the
parton distribution
\eqn\Mmom{ 
G(\xi,M)=\int^{\infty}_{-\infty}\! dt\, e^{-Mt} G(\xi,t).
} 
This derivation was originally
performed~\jar\ at LLx by assuming the common validity of eq.~\xevol\
and eq.~\tevol\ in the region where $Q^2$ and $\xi$ are both
large. However, it was more recently
realized~\refs{\afp,\sxres,\sxphen}
 that the solution of eq.~\xevol\ coincides generally
with that of eq.~\tevol,
up to higher twist corrections, provided only that the kernel of the
former is related to that of the latter 
by a `duality' relation, and boundary
conditions are suitably matched. This implies that 
the domains of validity of these
two equations are  in fact the same in perturbation theory, 
up to power--suppressed corrections.

The derivation of duality is simplest when
the coupling does not run, in which case the relation between the kernels of the two
equations is
\eqn\dual{
\chi[\as,\gamma(\as,N)]=N.
}  
This result is immediately found
by writing both eq.~\tevol\ and~\xevol\ and their solutions in terms of
the inverse Mellin transform of the parton distribution
\eqn\NMmom{
G(N,M)=\int^{\infty}_{0}\! d\xi\, e^{-N\xi} G(\xi,M)
=\int^{\infty}_{-\infty}\! dt\, e^{-Mt} G(N,t).
}
The inverse Mellin of the solution to
eq.~\xevol\ is
\eqn\simpole{
G(N,M)=\frac{F_0(M)}{N-\chi(M,\as)},
}
where $F_0(M)$ is an $N$--independent boundary condition. The
large $t$ behaviour of $G(N,t)$ is determined by the rightmost
singularity of $G(N,M)$~\simpole\ in the $M$--plane, while
the contributions of additional singularities further to the 
left are suppressed by powers of $Q^2$. In
particular, perturbative singularities are given 
by solutions of the equation $\chi(\as,M)=N$, while singularities of
the boundary condition $F_0(M)$ are nonperturbative. 

Expanding the denominator of the solution~\simpole\ about its
rightmost perturbative 
singularity, the solution eq.~\simpole\ is thus seen to coincide,
up to higher twist and nonperturbative terms, with 
the inverse Mellin of the standard solution to the
renormalization group eq.~\tevol, namely
\eqn\simpole{
G(N,M)=\frac{\tilde F_0(N)}{M-\gamma(N,\as)}
}
provided the duality equation~\dual\ holds, and the $t$--independent
boundary condition $\tilde F_0(N)$ to eq.~\tevol\ is related to
$F_0(M)$ by
\eqn\bcmatch{\tilde F_0(N)=-
{ F_0(\gamma(\as,N))\over\chi^\prime(\gamma(\as,N))}.} 
This establishes the perturbative equivalence (duality) of eqs.~\xevol\
and~\tevol, up to higher twist corrections.
Using the perturbative expansion of
$\chi$~\chiexp\ in the duality relation~\dual, 
we then find that
$\chi_0$ determines $\gamma_s(\as/N)$, while $\as \chi_1$ leads to 
$\as\gamma_{ss}(\as/N)$, where the combination
$\gamma_s(\as/N)+\as \gamma_{ss}(\as/N)$ correspond respectively to all terms 
of order $(\as/x) (\as \xi)^n$ and $\as(\as/x)(\as \xi)^n$
 in the splitting functions.

When one goes beyond LLx, i.e. beyond
the leading--order approximation for $\chi$, the  running of the 
coupling cannot be neglected, and this derivation must be
re--examined. Indeed, 
in $M$ space the usual running coupling $\as(t)$ becomes a
differential operator:  taking only the one-loop beta function into 
account 
\eqn\ashdef{
\ash = \frac{\as}{1-\beta_0 \as \smallfrac{d}{dM}}+\cdots,
}
where $\beta_0$ is the first
coefficient of the $\beta$-function (so 
$\beta=-\beta_0\as^2+\cdots$), 
with the obvious generalization to higher orders.
Hence, assuming the coupling to run in the usual way with $Q^2$, the 
$\xi$-evolution equation eq.~\xevol\  becomes~\Liprun
\eqn\xevolrun{
\frac{d}{d\xi}G(\xi,M)=\chi(\ash,M) G(\xi,M),
} 
where the derivatives with respect to $M$ act on everything to the right, and 
$\chi$ may be expanded as in eq.~\chiexp\ keeping the powers 
of $\ash$ on the left. Clearly, an operator--valued evolution kernel is a
potential threat to factorization. Furthermore, one may ask whether
eq.~\xevolrun\ is indeed the proper form of the running--coupling 
BFKL equation. Here  we  will  show
that, besides being intuitively appealing, the evolution 
eq.~\xevolrun\ has the property that its solutions are still
the same as those of a
renormalization--group equation eq.~\tevol, with a suitably matched
kernel, to all perturbative orders and up to power corrections.
This implies consistency of eq.~\xevolrun\ with standard
perturbative evolution, and in particular that factorization is preserved.
Therefore, eq.~\xevolrun\ can be viewed as an
alternative representation of the standard renormalization--group equation.

It is clear from eq.~\xevolrun\ that 
running coupling effects begin at NLLx. In fact, to N$^k$LLx it is
sufficient to retain the first $k$ terms in the expansion of 
$\ash$~\ashdef\ in powers of $\as$; the ensuing equation can then be
solved perturbatively. To NLLx one finds~\sxap\ that this solution is
again the same as that of a dual $t$--evolution equation~\tevol,
provided the fixed--coupling duality relation
eq.~\dual\ is modified by letting $\as\to\as(t)$, and then by
adding to $\gamma_{ss}$ an extra term~\ciafrun\  
$\Delta \gamma_{ss}$ 
proportional to $\beta_0$:
\eqn\deltag{
\Delta \gamma_{ss}(\smallfrac{\as}{N})=-\beta_0
\frac{\chi_0''(\gamma_{s})\chi_0(\gamma_{s})}
{2\chi_0'^2(\gamma_{s})}.   
}
Equivalently,  the duality relation
eq.~\dual\ can be formally preserved, provided that $\as\to\as(t)$ and 
the function $\chi$
used in it is no longer identified with the BFKL kernel, but rather given
by an `effective' $\chi$ function
\eqn\chiefdef{\eqalign{&\chi_{\rm eff}(\as,M)=\chi(\as,M) +\Delta
\chi(\as,M),\cr
&\quad \Delta \chi(\as,M)=\as^2 \Delta\chi_1(M)+\dots,\cr
}}
where $\chi(\as,M)$ is obtained letting 
$\ash\to\as$ in the kernel of eq.~\xevolrun,~\foot{Notice 
that, beyond leading order, this is not the same as
the fixed--coupling kernel of eq.~\xevol,
which refers to the unphysical case of a theory where
$\beta_0=0$ identically.}
and the correction term $\Delta \chi$
to
NLLx is given by
\eqn\deltachi{
\Delta \chi_{1}(M)=\beta_0
\frac{\chi_0''(M)\chi_0(M)}
{2\chi_0'(M)}.   
}
Also, the matching of boundary conditions is now given by a more complicated
relation than
eq.~\bcmatch, involving derivatives of $F_0$ and $\chi_0$; however, it
remains independent of $t$, and only involves $N$ and $\as$ (at the
initial scale).
We have further verified by explicit computation that if the perturbative
solution of eq.~\xevolrun\ with the running coupling eq.~\ashdef\ is
pursued up to N$^2$LLx these results remain valid: the boundary
condition remains $t$--independent, and duality can be preserved,
provided only a further $\Delta \chi_2$ term is included in
$\chi_{\rm eff}$ eq.~\chiefdef.

The problem with the perturbative approach is that the correction
terms which must be included in the effective
$\chi $ eq.~\chiefdef\ have an unphysical singularity: $\chi_0(M)$ has a 
minimum at $M=\half$, so the denominator of $\Delta \chi_{1}(M)$
vanishes, resulting in a simple pole in the NLLx correction
$\Delta\chi_1$ eq.~\deltachi\
at $M=\half$. The NNLLx correction $\Delta\chi_2$ turns out to have
a fourth--order pole, and in fact at each extra order three extra powers
of $(\chi'_0)^{-1}$ appear.
As explained in ref.~\sxap, this leads to a perturbative
instability: as a consequence of the singularity, 
the splitting function $\Delta P_{ss}$ eq.~\psasy\ associated with the
anomalous dimension $\Delta\gamma_{ss}$ behaves as
\eqn\splinst{{\Delta P_{ss}(\as,\xi)\over
P_s(\as,\xi)}\toinf{\xi}\left(\as\xi\right)^2.} 
In practice, the coefficient of this rise turns out to be small enough
that its effects are negligible in the HERA region~\sxphen. However,
the situation remains unsatisfactory from a theoretical point of
view: it would be better to extend the
perturbative proof of factorization and
duality to the case when the running of the
coupling is included to all orders, and then sum up these perturbatively
unstable terms. 
In the sequel, we will do this, and
show that the $M=\half$ singularity and the 
corresponding ones that appear at higher orders in
$\as \beta_0$ are an artifact of the expansion, are not present 
in the all--order solution, and can thus be resummed.

\newsec{Running Coupling Duality and Factorization}

In this section, we prove to all orders the perturbative duality and
factorization introduced in the previous section, and we discuss some
general properties of the all-order solution to the running coupling
BFKL equation, specifically its apparently unphysical behaviour in the
$\xi\to0$ limit.
We start from the running coupling $\xi$ evolution
equation~\xevolrun. We consider first the case in which the running is
included at one loop according to eq.~\ashdef, and the kernel $\chi$ is
linear in the operator $\ash$:
\eqn\chiphidef{\chi(\as,M)=\ash \varphi(\as, M),}
where $\varphi$ is a function of $M$ and the fixed coupling $\as$. 
This is of course the case
when the leading order form of $\chi$ is considered, in which case
$\varphi=\chi_0(M)$, but also more generally whenever 
terms of higher order
in the fixed coupling $\as$ are included in $\varphi$, as in the case of
the resummed leading order kernel $\tilde\chi_0$ previously discussed
by us~\refs{\sxres,\sxphen}.  
In the remainder of this section we will 
omit the explicit dependence of
$\varphi$ on $\as$, since it will
not play any role in the ensuing proof of factorization and duality. 
The generalization of the proof to 
higher orders in the running
coupling $\alpha_s(t)$ will be discussed at the end of the section. In
sect.~5 we will then show that, in order to perform a running coupling
resummation to NLLx, it is
sufficient to consider  a kernel $\chi$ of the form
of eq.~\chiphidef, and we will also discuss higher order generalizations.

With the kernel~\chiphidef, after
taking a second Mellin transform~\Nmom, eq.~\xevolrun\ becomes
the differential equation
\eqn\Mdiffeqn{
(1-\beta_0 \as \smallfrac{d}{dM}) N G(N,M) + F(M)
=\as\varphi(M)G(N,M)
}
The function $F(M)$ is the boundary condition, obtained
by acting with the operator $(1-\beta_0 \as
\smallfrac{d}{dM})$ on the boundary condition $-F_0(M)$ of
eq.~\simpole. The  perturbative solution~\sxap\  to eq.~\Mdiffeqn\ discussed
in the previous section  is an expansion in $\as$ at fixed $\as/N$
and it can be built
up by an iterative procedure:
at order $n$ the order $n-1$ expression for $G(N,M)$ is
inserted in the right--hand side. 
 Clearly, this iterative solution is a linear 
functional of $F(M)$. 

However, it is also easy to determine the general solution to  the differential
equation~\refs{\Liprun,\ColKwie}:
\eqn\gensol{
G(N,M)=H(N,M) + \int_{M_0}^M\!dM'\frac{H(N,M)}
{H(N,M')}~\frac{F(M')}{\beta_0 \as N},
}
where $H(N,M)$ is the solution of the homogeneous equation
\eqn\genthree{
H(N,M)=H(N,M_0)\exp{\left[\frac{M-M_0}{\beta_0 \as}
-\frac{1}{\beta_0 N}\int_{M_0}^M\!dM'\varphi(M')
\right]}
}
with an  arbitrary initial condition $H(N,M_0)$. The 
inhomogeneous term can be written as
\eqn\sol{
G(N,M)=\int_{M_0}^M\!dM' \exp{\left[\frac{M-M'}{\beta_0 \as}
-\frac{1}{\beta_0 N}
\int_{M'}^M\!dM''\varphi(M'')\right]}\frac{F(M')}{\beta_0 \as N},
}
and thus depends linearly on $F(M)$ while being independent of 
$H(N,M_0)$. 

Let us now see how the iterative solution can be recovered 
from the general solution, 
by expanding it in $\as$ at fixed $\as/N$: the iterative
solution corresponds to the inhomogeneous term in eq.~\sol, 
while the homogeneous solution in eq.~\genthree\ 
vanishes faster than any fixed perturbative order in $\beta_0$. 
To see how this works, change
the integration variable in eq.~\gensol\ from $M'$ to 
\eqn\ydef{
y(M,M')=-\frac{1}{\beta_0\as N}\int_{M'}^{M}
[N-\as\varphi(M'')]dM''
}
The inhomogeneous solution~\sol\ then becomes
\eqn\Gy{
G(N,M)=\int_0^{y(M,M_0)}  e^{-y}\frac{F(M'(y,M))}
{[N-\as\varphi(M'(y,M))]}dy,
}
where $M'$ is implicitly defined as a function $M'=M'(y,M)$ by eq.~\ydef.
The expansion of $M'(y,M)$ in powers of $\as$ is 
\eqn\expMprime{
M'=M+\frac{\beta_0\as N}{[N-\as\varphi(M)]}y
+O(\as^2).
}
Using this expansion to do the integral, and noting that  
$y\to\infty$ in the limit $\as\to 0$,  
one finds 
\eqn\pert{
G(N,M)=\frac{F(M)}{[N -\as \varphi(M)]} + \frac{\as
\beta_0}{[N - \as\varphi(M)]}\frac{d}{dM}\frac{F(M)}
{[N -\as \varphi(M)]} + \dots,
}
which is the iterative solution of ref.~\sxap, discussed in the
previous section. Furthermore, the homogeneous solution~\genthree\ is
manifestly proportional to $e^{-y}$ and therefore vanishes 
to all orders in the perturbative expansion.
Thus in the following we drop the homogeneous term 
and study only the inhomogeneous term~\sol, which reproduces the
perturbative solution.

Starting from $G(N,M)$ one can do either of the inverse 
Mellin transforms, and obtain
$G(\xi,M)$ or  $G(N,t)$. We consider $G(N,t)$ first,  
in order to prove  that it satisfies a renormalization--group
equation of the usual form, and  study the modification
induced by running--coupling
effects on the anomalous dimension and the duality equation which
relates it to the kernel $\chi$.
By taking the inverse $M$-Mellin transform we can write
$G(N,t)$ in the form
\eqn\solNt{
G(N,t)=\int_{c-i\infty}^{c+i\infty}\!\frac{dM}{2\pi i}
 \exp{\Big[Mt+\frac{M-M_0}{\beta_0 \as} - \frac{1}{\beta_0 N}
\int_{M_0}^M\!dM''\varphi(M'')\Big]}  {I(N,M)\over N},}
where
\eqn\Idef{I(N,M)
\equiv \int_{M_0}^{M}\!dM' 
\exp{\Big[\frac{M_0-M'}{\beta_0 \as}+\frac{1}{\beta_0 N}
\int_{M_0}^{M'}\!dM''\varphi(M'')\Big]} \frac{F(M')}
{\beta_0 \as }.
}
It is well--known~\refs{\ColKwie,\ciafrun} that evaluating the
$M$ integral by saddle point one recovers a running--coupling version
of the duality relation eq.~\dual. However, both duality and
perturbative factorization seem to be spoiled in the process. We now
review this
derivation, and then prove that, contrary to appearances, 
factorization still holds in the
perturbative limit, and duality is recovered with a series of
corrections eq.~\chiefdef.

The saddle--point condition for the integrand of eq.~\solNt\ 
(assuming $I(N,M)$ to be a smooth function of $M$) is
\eqn\Ms{
t + \frac{1}{\beta_0 \as} - \frac{1}{\beta_0  N}\varphi(M_s)
=0.
}
Noting that at leading order 
$t+1/(\beta_0 \as)= \left[\beta_0\alpha_s(t)\right]^{-1}$, 
we see that this is equivalent to 
\eqn\Msone{
\alpha_s(t)\varphi(M_s)=N.
}
Identifying  $M_s$ with the anomalous dimension, 
\eqn\lsadc{M_s=\gamma_s(\alpha_s(t)/N),} 
we see that eq.~\Msone\
coincides with the duality relation, eq.~\dual, but
with the 
fixed coupling replaced by the running coupling~\refs{\ColKwie,\ciafrun}. 
Substituting back the saddle condition in the exponent of eq.~\solNt\
we get (differentiating the duality relation with respect to $t$)
\eqn\gam{
\frac{M_s}{\alpha_s(t)\beta_0}-\frac{M_0}{\alpha_s\beta_0} - \frac{1}{\beta_0\as N}
\int_{M_0}^{M_s}\!dM'\varphi(M')
=\int_{t_0}^t\! dt' \gamma_s(\alpha_s(t')/N),
}
where $\gamma_s(\alpha_s(t')/N)$ is defined for all $\as(t')$
as the solution of eq.~\lsadc, and
we choose for simplicity (and with no loss of generality)
$\as$ such that $\alpha(t_0)=\as$. Performing the saddle 
integral we thus find
\eqn\solNtone{
G(N,t)=\sqrt{\frac{N\beta_0\as}{-\varphi'[M_s(t)]}}
 \exp{\left(\int_{t_0}^t\!dt'\gamma_s(\alpha(t'),N)\right)}
\frac{I(N,M_s(t))}{N}
+\cdots,
}
where the dots denote corrections to the leading saddle--point approximation.

Eq.~\solNtone\ seems to violate factorization because of the
$t$--dependence induced
by the substitution of the saddle
condition $M=M_s(t)$ in the integral $I(N,M)$, which depends on the
boundary condition. Also, it seems to  violate duality because  the
$t$--dependent prefactor $1/\sqrt{\varphi'}$ spoils the identification
of the anomalous dimensions with $\gamma_s$.
This is puzzling because these violations seem to
start at NLLx, whereas the explicit perturbative calculation discussed
above shows that factorization and duality are  respected at NLLx.
We now address both issues in turn.

First, let us consider the integral $I(N,M_s(t))$ 
which appears in eq.~\solNtone (with $I(N,M)$ defined in eq.~\Idef\ and
$M_s$ eq.~\lsadc)
in the
perturbative LLx and leading--log~$Q^2$ limit, i.e. as $\as \rightarrow 0$, with $\as/N$ 
and $\as t$ fixed. We change the integration variable from $M'$ to 
$y(M_0,M')$ defined in eq.~\ydef, so  we can write this integral in the form
\eqn\Ilim{
I(N,M_s(t))=\int_0^{y[\gamma(\alpha(t)/N)]}\!dy e^{-y}
 \frac{F(M'(y,M_0))}{[N - \varphi(M'(y,M_0))]}
}
We take $(1-\frac{\as}{N} \varphi)<0$ 
and $M'<M_0$, so that $y>0$. 
As a result, 
in the perturbative limit, $y(\gamma) \rightarrow \infty$ and the perturbative $t$
dependence disappears to all orders in the expansion. So 
perturbatively $I(N,M_s(t))$ is independent of $t$  to all orders, even
though it does depend on $t_0$ because $M'=M'(y,M_0)$.
This  proves factorization to all perturbative orders: the $t$
dependence of $G(N,t)$ is entirely determined by the evolution kernel
$\chi$, and does not depend on the boundary condition

The $t$--dependence, however, is not only due to the anomalous
dimensions $\gamma_s(\as(t)/N)$ which satisfies the running--coupling
duality relation eqs.~\Msone,\lsadc, but also to the square--root
prefactor in eq.~\deltag.
However, because this  additional $t$--dependence goes entirely
through the running coupling $\alpha_s(t)$, it has the structure of a NLLx
coefficient function, and it can thus
be reabsorbed by small--$x$ scheme change~\refs{\sxap,\sxphen,\mom} 
in a redefinition of the 
anomalous dimension and the boundary condition:
\eqn\deltagam{\eqalign{
\frac{1}{\sqrt{-\varphi'(\gamma_s(\smallfrac{\alpha_s(t)}{N}))}}
&=\exp-\left[\half\ln{-\varphi'
(\gamma_s(\smallfrac{\alpha_s(t)}{N}))}\right]\cr
&=\exp\left[\beta_0\int_{t_0}^t\!dt'\,
\frac{\varphi''(\smallfrac{\alpha_s(t')}{N}) 
\varphi(\smallfrac{\alpha_s(t')}{N})}
{2\varphi'^2(\smallfrac{\alpha_s(t')}{N})} \as(t') 
\right] \frac{1}{\sqrt{-\varphi'(\gamma_s(\smallfrac{\alpha_s(t_0)}{N}))}}\cr
&
=\exp\left[\int_{t_0}^t\!dt'\,\Delta \gamma_{ss}(\as(t')/N)
\right] \frac{1}{\sqrt{-\varphi'(\gamma_s(\smallfrac{\alpha_s(t_0)}{N}))}},  
}}
where the  second equality follows from differentiating
the duality relation, and  the last factor, which depends on the
initial scale $t_0$
only, can be reabsorbed in the boundary condition.
Identifying  $\varphi=\chi_0$ we see immediately that
the resulting
addition to $\gamma$,
$\Delta \gamma_{ss}$,
is the same as that determined perturbatively at NLLx
eq.~\deltag. 

Furthermore, we can use the saddle--point method to generate an
asymptotic expansion of $G(N,t)$. Noting that the first correction to the
gaussian approximation is a cubic term in $M-M_s$ proportional to
$\varphi^\prime$, and that odd terms in the expansion
vanish upon integration, this is seen to
generate an expansion in powers of
\eqn\sadexp{
{N\beta_0\over{\varphi'}^3}=\as\left({\as^2\over
N^2}\right)^2[\gamma_s'(\as(t)/N)]^3
\beta_0,}
where $\gamma_s'$ is differentiated with respect to its argument
$\as(t)/N$. It follows that higher orders in the expansion give a
series of terms which can be absorbed in a series of contributions to
the anomalous dimension of the form
\eqn\gamexpone{
\gamma_{\rm rc}(\alpha_s(t),N)
=\gamma_s(\smallfrac{\alpha_s(t)}{N})
+\beta_0\alpha_s(t)\Delta \gamma_{ss}(\smallfrac{\alpha_s(t)}{N})+
[\beta_0\alpha_s(t)]^2\Delta \gamma_{sss}(\smallfrac{\alpha_s(t)}{N})+\dots.
}
where $\Delta\gamma_{ss}$, given by eq.~\deltag, is due to the leading
fluctuations about the saddle, $\Delta\gamma_{sss}$ comes from
the next--to--leading fluctuations and so on. It is also easy to check that the
boundary condition does not contribute to this additional $t$
dependence, because all its higher order derivatives with respect to
$M$ vanish in the perturbative limit, for the same reason that the
homogeneous term vanishes to all perturbative orders.

This shows that we can formally preserve the duality relation
eq.~\dual, with $\as\to\as(t)$ and an effective $\chi_{\rm eff}$
eq.~\chiefdef, 
provided  at each order in $\alpha_s$ 
we add a new correction term to the kernel $\chi$, determined from
$\gamma_{\rm rc}$ eq.~\gamexpone; 
the first correction is  given by eq.~\deltachi.
The order $\alpha_s^k$ correction is proportional to 
$[\varphi']^{-3k}$ eq.~\sadexp, and thus if $\varphi$ has a minimum, 
the correction  has a $3k$-th order pole as
$M\to M_{\rm min}$. Therefore~\sxap, when $\xi\to\infty$ the 
corresponding splitting function 
rises by a factor $\xi^{2k}$
faster than the leading order, and the perturbative expansion is unstable.
These singular contributions will be resummed in sect.~4--5, leading
to an anomalous dimension $\gamma$ and associated kernel $\chi_{\rm
eff}$  which are free of singularities, and thus have stable
perturbative expansions.

Before we turn to this task, let us exploit further the factorization
property which follows from
the fact that the integral $I(N,M_s(t))$ 
in eq.~\Ilim\ is asymptotically a function of
$N$, independent of $t$ to all orders in the perturbative 
expansion in $\alpha_s(t)$. This implies that, in the same
limit, $G(N,t)$ in eq.~\solNtone\ can be written as
\eqn\xthree{\eqalign{
G(N,t)&= \int_{c-i\infty}^{c+i\infty}\!\frac{dM}{2\pi i} e^{Mt}\cr
 &\qquad\exp{\Big[\frac{M-M_0}{\beta_0 \as}
 - \frac{1}{\beta_0
N}\int_{M_0}^M\!dM''\varphi(M'')\Big]}\frac{I(N,t_0)}{N}.
}}
Note that this is the same form that one could obtain from $G(N,M)$ 
given by the homogeneous solution in eq.~\genthree\ 
with the identification $H(N,M_0)=I(N,t_0)/N$, that is with an initial 
condition which is a linear functional of $F(M)$.  

This proof of factorization and duality can be extended to the case in
which the running of the coupling is included to higher orders.
To this purpose, introduce a perturbative
expansion of the inverse $M$-Mellin $G(\xi,t)$ of the
solution to eq.~\xevolrun\ in powers of $\as(t)$:
\eqn\rcgexp{G(\xi,t)=G_0(\xi,t)+ \as(t) G_1(\xi,t)+\dots.} 
Then, the double Mellin  $G_0(N,M)$ satisfies
eq.~\Mdiffeqn\ with $\varphi=\chi_0$, while $G_1$ satisfies
\eqn\nloeq{NG_1(N,M)=\ash \left[\chi_0(M) G_1(N,M)+ \chi_1(M)
G_0(N,M)\right]+O(\ash^2),}
where we have made use of the fact that the commutator of $\ash$ and
$\chi_0$ is of order $\ash^2$.
Eq.~\nloeq\ is immediately recognized to have the same form
as eq.~\Mdiffeqn, with a boundary condition now given by the $\chi_1
G_0$ term. Hence, its solution has the form eq.~\gensol\ and
factorization and duality are preserved. Specifically, $G_1$
is obtained multiplying the factorized $G_0$ eq.~\xthree\ by a suitable
coefficient determined by $\chi_1$, and thus  it has
the structure of the standard perturbative solution, 
but now with the factor of
$\as$ which multiplies $\chi_1$ promoted to running
$\as(t)$.
The procedure can then be pursued to higher orders: in fact, thanks to
the expansion of $G$ in powers of $\as(t)$, 
the higher--order differential equation for $G$ which is
found by simply letting $\as\to\ash$ in the higher orders of the
expansion of $\chi$ eq.~\chiexp\ is reduced to
a linear differential equation for each individual $G_i$.

Finally, it is interesting to observe that the inverse
$N$--Mellin transform of $G(N,M)$ can actually be evaluated exactly
if the boundary condition $I(N,M)$
is assumed to be a sufficiently smooth function of $N$, so  that it may be
effectively considered to be $N$--independent, i.e. assuming that the
boundary condition is $x$--independent, and all further $x$ dependence
is generated by perturbative evolution. Indeed, using for definiteness
eq.~\xthree\ and setting
$I(N,t_0)= G_{00}$ one gets~\Haak
\eqn\genfour{
G(\xi,M)=G_{00}\exp{\frac{M-M_0}{\beta_0 \as}}  
I_0\left(2\sqrt{\frac{\xi}{\beta_0}
\int_{M_0}^M\!dM'\varphi(M')}\right),
}
where $I_0$ is the modified Bessel function. 
Even though it is not possible to determine the inverse $M$--Mellin
transform of eq.~\genfour\ in closed form, the asymptotic behaviour of
$G(\xi,t)$ when $t$ is large may be determined by performing the
inverse $M$--Mellin
by saddle point. The saddle condition is
\eqn\gensix{
\sqrt{\frac{\beta_0}{\xi}} \frac{1}{\beta_0
\as(t)}=\frac{I'_0}{I_0} \frac{\varphi(M)}
{\sqrt{\int_{M_0}^M\!dM'\varphi(M')}}.
}
In the limit of large $\xi$, the left--hand side of eq.~\gensix\
vanishes, while on the right--hand side 
$I'_0/I_0\toinf{\xi}1$ so  the whole expression vanishes
only if $\varphi$ goes through zero for some $M$  (recall that we
assume
$M\le M_0$ in order for the perturbative limit to exist). Clearly if 
$\varphi$ is always positive, as is the case for $\varphi$ which has a
minimum with a positive value, the saddle equation cannot be 
satisfied at a real value of $M$: in such case, the asymptotic
behaviour   of $G(\xi,t)$ is dominated by a pair of complex saddle
points, and therefore displays unphysical
oscillations. Alternatively, if 
$\varphi$ has a vanishing or negative minimum, there is always a real saddle.
The asymptotic limit is then stable, provided we choose $M_0$ equal to
the value of $M$ where $\chi$ vanishes (otherwise, for negative
minimum the square--root in eq.~\genfour\ becomes complex).

The onset of unphysical behaviour at large enough $\xi$
has been interpreted as a signal of breakdown of perturbation
theory~\refs{\ciaf,\ColKwie,\Bartels\Thorne{--}\CiaMue}, and thus it
has been viewed as an indication that the perturbative approach
must be abandoned in the small $x$ limit. However, 
in the following section we will 
show that if we
are interested in the BFKL approach as a tool to improve the 
theory of scaling violations at small $x$, then this negative
conclusion is unjustified. Indeed,
even when $G(\xi,t)$ determined by inverse Mellin from
eq.~\genfour\
displays an asymptotic oscillatory behaviour, the associated 
splitting function remains smooth and monotonic, and therefore no
oscillations are seen in the factorized solutions to leading--twist
evolution equations.  

\newsec{Quadratic Kernel and Asymptotic Behaviour}

In order to resum the running--coupling corrections to duality
eq.~\gamexpone\ we must determine the parton distribution
 $G(N,t)$ by performing the inverse $M$-- Mellin transform
in eq.~\solNt\ exactly,
rather than in the saddle-point approximation as in the
previous section. Even though it is nontrivial to do this in general,
the integral 
can be easily determined~\refs{\Liprun,\ciafrun}
if the kernel $\varphi$ in eq.~\solNt\ is quadratic. The ensuing
form of the parton distribution displays the unphysical oscillations
discussed at the end of the previous section. 
In this section we review this determination of $G(N,t)$, and then we 
use it to determine the running--coupling dual anomalous dimension
eq.~\gamexpone, and show that  the corresponding splitting
function, and thus the solution to the associated evolution equation,
are free of oscillations and have a stable asymptotic $\xi\to\infty$
limit. In the next section we will then show that knowledge of the
resummed anomalous dimension for a quadratic $\chi$ is sufficient to
resum running--coupling singularities in general.

We start assuming that $\varphi$ is given by
\eqn\aione{
\varphi(M)=c + {1\over 2} \chipp (M-\half)^2,
}
where $\chipp$ and $c$ are constants. 
In the sequel we will discuss the three cases of positive, negative,
or vanishing $c$. While we will assume that $\kappa$ is positive but
otherwise generic, for explicit numerical examples we will take
\eqn\kapval{\kappa=-\smallfrac{2 C_A}{\pi}\psi''(\half),} 
(where $\psi(x)$ is the digamma function and $C_A=3$)
which corresponds to the
curvature of the leading--order BFKL kernel~\bfkl\ $\chi_0$~\chiexp\
around its minimum.
The expression of 
$G(N,t)$ is then  given in terms of an Airy
function~\Liprun:  inserting 
eq.~\aione\ in eq.~\xthree\  
and performing  the $M$ integral with the choice
$M_0=\half$, we get
\eqn\airysoln{
G(N,t)=K(N)  e^{{1/[2\beta_0\alpha_s(t)]}}
\Ai[z(\as(t),N)]\frac{I(N,t_0)}{N} ,
}
where $\Ai(z)$ is the Airy function, which satisfies
\eqn\airy{
\Ai''(z)-z\Ai(z)=0,
}
with $\Ai(0)=3^{-2/3}/\Gamma(2/3)$, 
\eqn\aitwo{
z(\as(t),N) \equiv 
\left(\frac{2\beta_0 N}{\chipp}\right)^{1/3} \frac{1}{\beta_0}
\left[\frac{1}{\alpha_s(t)}-\frac{c}{N}\right],
}
and $K(N)$ is a $t$-independent normalization,
\eqn\kdef{K(N)= e^{-\left[{1\over2\beta_0\as}\right]} \left(\frac{2\beta_0
N}{\chipp}\right)^{1/3} {1\over\pi}.}

Along the positive real axis, the Airy function is a positive
definite, monotonically decreasing function of its
argument, and it behaves asymptotically as
\eqn\airyasymp{
\Ai(z)= \half \pi^{-1/2}z^{-1/4}
\exp(-\smallfrac{2}{3}z^{3/2})\big(1+O(z^{-3/2})\big).
}
For large enough $N$, $z\sim N^{1/3}$, so $\Ai[z(\as,N)]$ is also a
monotonically decreasing function of $N$, asymptotically 
damped as $\Ai[z(\as,N)]\sim \exp -\sqrt{N}$.
When $c=0$, this monotonic behaviour persists for all real
positive $N$ even when $N$ is small. However, if $c>0$, $z$ changes sign when 
$N<\alpha_s(t)c$, and $\Ai(z)$ oscillates
along the negative real $z$ axis, so  $\Ai[z(\as,N)]$ oscillates 
as $N\to 0$, because $z\to-\infty$ in this limit. 
If instead $c<0$, $z\to\infty$ not only as $N\to\infty$ 
but also as $N\to0$, and  $\Ai[z(\as,N)]$ is smoothly damped in both limits.

\topinsert
\vbox{
\epsfxsize=7.5truecm
\centerline{\epsfbox{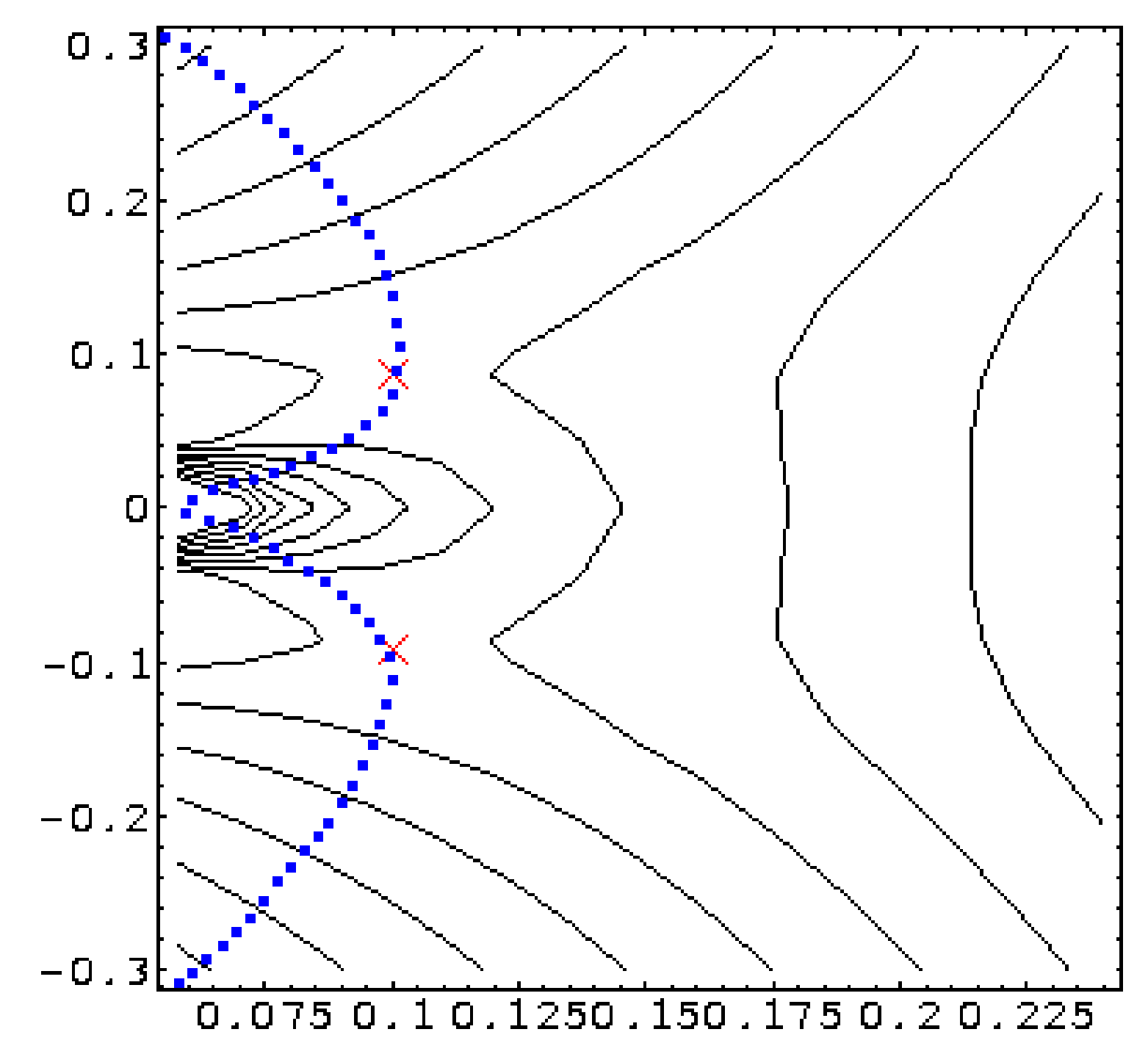}}
\hbox{
\vbox{\footnotefont\baselineskip6pt\narrower\noindent
Figure 1: Contour plot in the complex $N$ plane for the Mellin
inversion integral of $G(N,t)$. The crosses indicate the approximate location
of the complex saddles, and the dotted line is the approximate
steepest descent path.
}}\hskip1truecm}
\endinsert
It is now easy to check that the 
behaviour of $G(\xi,t)$ obtained by inverse Mellin transformation from
$G(N,t)$~\airysoln\  has the asymptotic properties discussed in
general at the
end of sect.~3. Specifically,
if $c$ (i.e. the value of $\chi$~\aione\ at its minimum) is 
positive, then $G(\xi,t)$ displays an unphysical oscillatory behaviour 
as $\xi\to\infty$.  This can be seen explicitly by evaluating by saddle
point the $N$--Mellin
inversion integral which gives  $G(N,t)$. The saddle point
condition is dominated by the Airy
function, and has thus the form
\eqn\aitwoe{
\xi=- {d\Ai[z(\as,N)]/dN\over\Ai[z(\as,N)] }.
}
If $\xi$ is not too large, the saddle condition is satisfied in the large
$N$ region where $\Ai[z(\as,N)]$ is a decreasing function of $N$. As
$\xi$ grows, the saddle is drawn towards $N\to0$ where $\Ai[z(\as,N)]$
decreases faster with $N$; however, for
sufficiently small $N$ the derivative changes sign and the real saddle
is lost. This situation is displayed in
fig.~1, where we 
show a contour plot in the complex $N$-plane
of the real part of the integrand to the $N$--Mellin inversion,
namely  $\Re\left[\xi N+\ln G(N,t)\right]$ with $G(N,t)$
given by eq.~\airysoln,  
$I(N,t_0)=1$, and $t=6$ (corresponding to
$\as(t)\approx 0.2$),  $\xi=10$, $c=1$. 
Along the real axis the function increases, and there is no saddle.
However, even when $\xi\to\infty$ the saddle condition eq.~\aitwoe\ does
still
have a pair of complex
solutions, with $|N|\to0$ as $\xi$ grows. These saddle points 
can be clearly seen in fig.~1.

For large $\xi$, the position of the complex 
saddle points can be determined explicitly using the asymptotic
expansion eq.~\airyasymp, which is valid when $|\arg(z)|<\pi$: one
finds 
\eqn\saddles{
N_\pm = e^{\pm i\pi/4} (k/\xi)^{1/2},
}
where $k\equiv \smallfrac{2}{3}\smallfrac{c}{\beta_0}
(\smallfrac{2c}{\chipp})^{1/2}$.
The asymptotic large $\xi$ behaviour is dominated by the contributions
from this pair of complex conjugate saddle points, which add 
to give cosine oscillations: explicitly 
\eqn\oscillations{
G(\xi,t) \toinf{\xi} 
\exp(\sqrt{2k\xi})
\cos\left(\sqrt{2k\xi}
\right).
}
As $\xi$ becomes large, $G(\xi,t)$ becomes negative and starts 
to oscillate, so the solution looks unphysical.

If $c=0$, it is instead easy to see that eq.~\aitwoe\ has a solution
for all $\xi$, with $N\to0$ as $\xi\to\infty$. If, on the contrary,   $c<0$ the
real saddle is again lost as $N$ decreases because of the change of
sign in the derivative of $dz/dN$. However, if one chooses
$M_0\not=\half$ the solution eq.~\airysoln\ acquires an extra
$N$--dependent prefactor, and
it can be shown that 
one recovers a stable
asymptotic behaviour if $M_0$ is chosen so that
$\varphi(M_0)=0$. Hence, also in these cases one recovers the generic
behaviours found from the Bessel function representation of
$G(\xi,M)$ eq.~\genfour.

At first sight the unphysical oscillatory behaviour of $G(N,t)$ when
$c>0$ appears 
disastrous~\refs{\ciaf,\Liprun\ColKwie\KovMue\Bartels\Thorne{--}\CiaMue}.
However, here we are not directly interested in the solution
eq.~\airysoln ,
but rather in the dual anomalous dimension which can be extracted from
it.
This is given by
\eqn\aithree{
\gamma_A(\as(t),N)=\frac{d}{dt}\ln G(N,t)=\frac{1}{2}
 + \left(\frac{2\beta_0
N}{\chipp}\right)^{1/3}\frac{\Ai'[z(\as(t),N)]}{\Ai[z(\as(t),N)]},
}
which we now study.
The large--$N$ behaviour of $\gamma_A$ is easily determined, because
when $N$ is large, $z$ is large and we can use the
asymptotic expansion eq.~\airyasymp\ in eq.~\aithree. This gives
\eqn\aifive{
\gamma_A(\as,N) \sim \frac{1}{2} - \sqrt{\frac{2}{\chipp}
\Big[\frac{N}{\as}-c\Big]}+O\left({1\over [z(\as,N)]^{3/2}}\right).
}  
Note that this leading behaviour coincides with the `naive dual' anomalous
dimension: namely, the anomalous dimension 
which is found from $\varphi$ eq.~\aione\ using
the leading--order duality relation~\dual, but with $\as=\as(t)$. We will
come back to this point in the next section.
\topinsert
\vbox{
\epsfxsize=9.8truecm
\centerline{\epsfbox{n0.ps}}
\hbox{
\vbox{\footnotefont\baselineskip6pt\narrower\noindent
Figure 2: Location of the rightmost singularity of the Airy anomalous
dimension~\aithree, $N_0(t)$   (solid) and of the naive dual anomalous
dimension~\aifive, $\as(t) c$ (dotted).
 In both cases, the curves correspond (top to bottom) to
$c=2,\,1.5,\,1,\,0.5$. 
}}\hskip1truecm}
\endinsert

In order to determine the small--$N$ behaviour of $\gamma_A$
we must distinguish three cases.
For positive 
values of $c$, as $N$ decreases $\gamma_A(\as(t),N)$ 
increases indefinitely until it blows up for the
value $N=N_0(t)$ that corresponds  
to the first zero $z_0\approx-2.338\dots$ of $\Ai(z)$. Because
$\Ai'$ is regular at $z=z_0$, at $N=N_0$ the anomalous dimension has a
simple pole. The location of
this rightmost singularity of the anomalous dimension 
can be determined by solving the implicit equation
$z[\as(t),N_0]=z_0$, with $z[\as(t),N_0]$ given by eq.~\aitwo. For
small values of $\as(t)$ this solution can be expanded as
\eqn\aitwob{
N_0(\as,z)=c\alpha_s[1 + \sigma + \frac{2}{3}\sigma^2
 + \frac{1}{3}\sigma^3 + O(\sigma^4)],
}
where
\eqn\aitwoa{
\sigma(\as,z) \equiv (\beta_0\alpha_s)^{2/3}
\left(\frac{\chipp}{2c}\right)^{1/3}.
}
Recalling that $z\to-\infty$ as $N\to0$ and  $z_0<0$,
eq.~\aitwob\  implies that $N_0<\as c$.
A plot of the position of the rightmost singularity 
$N_0(t)$ for several values of $c$ is displayed in fig.~2, along with
the position of the rightmost singularity of the naive dual anomalous
dimension eq.~\aifive, namely the square--root branch cut at $N=\as(t)
c$.~\foot{In this and subsequent figures we take the standard two-loop form
of $\as(t)$, with $\as(M_z)=0.119$.}
Below $N_0$, as $N\to0$, $z$ goes through an infinite number of zeros
of the Airy function, and the
anomalous dimension oscillates wildly. A plot of the anomalous
dimension along the positive real axis (with $c=1$)  is displayed in 
fig.~3: at small $N$ the anomalous dimension goes through an infinity
of poles as $N\to0$, while at large N the onset 
of the square--root drop eq.~\aifive\ is seen.
\topinsert
\vbox{
\epsfxsize=9.8truecm
\centerline{\epsfbox{aiad.ps}}
\hbox{
\vbox{\footnotefont\baselineskip6pt\narrower\noindent
Figure 3: The Airy anomalous dimension $\gamma_A(\as,N)$ 
for $c=1$ and $\as=0.2$ ,
showing the oscillations below the Airy zero at $N_0(t)$.
}}\hskip1truecm}
\endinsert

If $c\leq0$, it is easy to see that $\Ai[z(\as(t),N)]$ does not have
any zeros in the complex $N$ plane. Indeed, the zeros of $\Ai(z)$ 
are all located on the negative real $z$--axis. However, 
the cut induced by the function $z(\as(t),N)$~\aitwo\ must be taken
along the negative $N$--axis, so that the
$N$--Mellin inversion integral is well--defined in the physical region
$N>0$. But the function $z(\as(t),N)$ then maps the cut $N$ plane on
the quadrant $-{2\pi\over 3} \leq z \leq {2\pi\over 3}$ (if $c<0$) 
or $-{\pi\over 3} \leq z \leq {\pi\over 3}$ (if $c=0$). 
In either case,
all the zeros of the Airy function lie outside the physical sheet of
the $N$ plane. 
Furthermore, in both cases the anomalous dimension
eq.~\aifive\ turns out to have a cut along the negative $N$ axis starting at $N=0$,
which determines the small $x$ asymptotics.

Specifically, if $c=0$, then  $z\propto N^{1/3}$ for all $N$. Because 
$\smallfrac{\Ai'(0)}{\Ai(0)}=- 3^{1/3}
\smallfrac{\Gamma(2/3)}{\Gamma(1/3)}$ the anomalous dimension has a
cubic root branch cut at $N=0$ and is regular for $N>0$.
If instead $c<0$, the cut can be seen
 by using the
representation of the Airy function in terms of the modified
Bessel functions $K_\nu$
to rewrite the anomalous dimension~\aifive\ as
\eqn\abes{\gamma_A(\alpha_s,N)=
\left(\frac{2(N-\as c)}{\kappa\as}\right)^{1/2} 
{K_{2/3}(\smallfrac{2}{3} z^{3/2})\over K_{1/3}(\smallfrac{2}{3} z^{3/2})}.}
It is then easy to show using the continuation formula
\eqn\con{K_\nu(-x) = e^{-i\nu\pi}K_\nu(x)-i\pi I_\nu(x)}
that $\gamma_A$ has a discontinuity along the
negative real $N$ axis,  given by (as $|N|\to 0$) 
\eqn\disc{{\rm disc}(\gamma_A)=-{2i} \left({-2
c\over\kappa}\right)^{1/2} \exp- \left[\frac{4}{3}
{1\over \beta_0 |N|}\sqrt{\frac{-2c^3}{\kappa}}\right]+O(N).}
Notice that if
$c<0$, then  when $N$ is small, $z$ is 
large and the expansion eq.~\aifive\ holds. This would seem
to suggest that  the anomalous 
dimension is regular and positive at $N=0$. However, this conclusion
cannot be drawn because this expansion is merely asymptotic: in fact,
it is factorially divergent, though Borel summable. Its Borel sum
coincdes with eq.~\abes, with the discontinuity eq.~\disc, which is
exponentially subleading and thus not immediately visible in the
asymptotic expansion eq.~\aifive.

\topinsert
\vbox{
\epsfxsize=9.8truecm
\centerline{\epsfbox{splfun.ps}}
\hskip4truecm\hbox{
\vbox{\footnotefont\baselineskip6pt\narrower\noindent
Figure 4: The Airy splitting function $P_A(\as,x)$ 
with $\as=0.2$, and (top to bottom) $c=2,\>1,\>0,\>-1,\>-2$,
showing the absence of oscillations at very small $x$ (solid). The
standard two--loop splitting function is also shown (dashed).
}}\hskip1truecm}
\endinsert
We can now study the splitting function $P_A$, related by Mellin transform
eq.~\psasy\ to the Airy anomalous dimension $\gamma_A(\as(t),N)$.
If $c>0$, the asymptotic $\xi\to\infty$ behaviour of the splitting
function is dominated by the pole at $N=N_0(t)$, which is the
rightmost singularity of the anomalous
dimension: 
\eqn\airysplit{
x P_A(\as(t),x)\tozero{x}\as(t)
x^{-N_0(t)}.
}

If $c\leq0$, the small $x$  behaviour is instead controlled by the cut
along the negative $N$ axis. Specifically, if $c=0$ branch cut is
cubic, and the asymptotic behaviour
can be estimated noting that the inverse Mellin of
$N^{1/3}$ is proportional to a gamma function: consequently when $c=0$
\eqn\splitas{
xP_A(\as(t),x)\tozero{x}
{1\over \xi^{4/3}}.}
If instead $c<0$ the asymptotic behaviour can be estimated by
evaluating by saddle point the integral along the discontinuity
eq.~\disc, with the result
\eqn\splitasneg{
xP_A(\as(t),x)\tozero{x}
\exp\left[ - \frac{4}{\sqrt{3}}
\left(\frac{-2c^3}{\kappa}\right)^{1/4}
\left({\xi\over\beta_0}\right)^{1/2}
\right].}

Therefore,  asymptotically
if $c>0$ the splitting function grows as an inverse power
of $x$, if $c=0$ it drops as a power of $\xi=\ln (1/x)$, and if $c<0$
it drops faster than any power of $\xi$ but slower than a power of $x$.
Whatever the value of $c$, 
the splitting function is smooth for all $\xi>0$. 
A plot of $x P_A(\as(t),x)$ for $c=2,\>1,\>0,\>-1,\>-2$
is displayed in
fig.~4, along with the usual two-loop splitting function:
it is apparent that
the splitting function is well--defined and 
positive for arbitrarily small $x$. The fact that the anomalous
dimension has singularities for $N\le N_0(t)$ (if $c>0$) only means that
$N_0(t)$ is the lowest moment 
of the splitting function for which the integral eq.~\psasy\  
over $x$ converges, but does not imply that there are problems 
in the $t$-evolution. The effect
of the running of the coupling is to replace the naive dual anomalous
dimension eq.~\aifive\ with the resummed anomalous dimension
eq.~\aithree. We can then equivalently
view the effect of the running of the coupling
as a modification of the evolution
kernel, whereby the duality relation~\dual\ is preserved, but the
`bare' kernel $\varphi(M)$~\aione\ is replaced by a resummed Airy effective
kernel, defined by the duality equation
\eqn\efker{\as\chi_A(\gamma_A(\as,N))=N}
in terms of the Airy anomalous dimension
$\gamma_A(\as,N)$~\aithree.
In the next section we will
show explicitly that the perturbative effective
kernel eq.~\chiefdef\ is an (asymptotic) expansion of the Airy effective
kernel eq.~\efker.
The bare and effective kernel (when $c=1$) are
compared in fig.~5. Note that indeed at small $M$ (i.e., by duality,
large $N$) the resummed and naive kernel coincide, up to a subleading 
(asymptotically constant) correction, in agreement with eq.~\aifive. 
\topinsert
\vbox{
\epsfxsize=9.8truecm
\centerline{\epsfbox{chiquad.ps}}
\hskip4truecm\hbox{
\vbox{\footnotefont\baselineskip6pt\narrower\noindent
Figure 5: The kernel $\varphi(M)$ eq.~\aione\ (dashed) and the
Airy effective kernel $\chi_A(M)$ eq.~\efker\  (solid)
with $c=1$ and $\as=0.2$.
}}\hskip1truecm}
\endinsert

Because the resummed splitting function is smooth, the solution to 
perturbative evolution driven
by it is also smooth and free of
oscillations. This can be understood by noting that the solution of
the evolution equation~\tevol\ with the 
anomalous dimension eq.~\aithree\ and the boundary condition 
$G(N,t_0)=\tilde F_0(N)$ is
\eqn\aitwof{\eqalign{
G(N,t)&=\Gamma(t,t_0;N)
\tilde F_0(N),\cr
\Gamma(t,t_0;N)&\equiv e^{\int_{t_0}^t\gamma_A(\as(t),N)dt}=
{G(N,t)\over G(N,t_0)},\cr}}
with the evolution factor $\Gamma(t,t_0;N)$ explicitly given by
\eqn\evfact{\Gamma(t,t_0;N)=
e^{1/[2\beta_0\alpha_s(t)]-1/[2\beta_0\alpha_s(t_0)]}
{\Ai[z(\as(t),N)]\over \Ai[z(\as(t_0),N)]}.}
Now,  for large enough $\xi$, the oscillations 
eq.~\oscillations\ are the same at all values of $t$, and thus cancel
in the ratio $G(N,t)/G(N,t_0)$. 
This is a consequence of the fact that
the saddle points which drive the large $\xi$ oscillations are  
dominated by the $t$--independent  $-c/N$ contribution to $z$~\aitwo, which
at small $N$ overwhelms the $1/\alpha_s(t)$ term.

\topinsert
\vbox{
\epsfxsize=7.5truecm
\centerline{\epsfbox{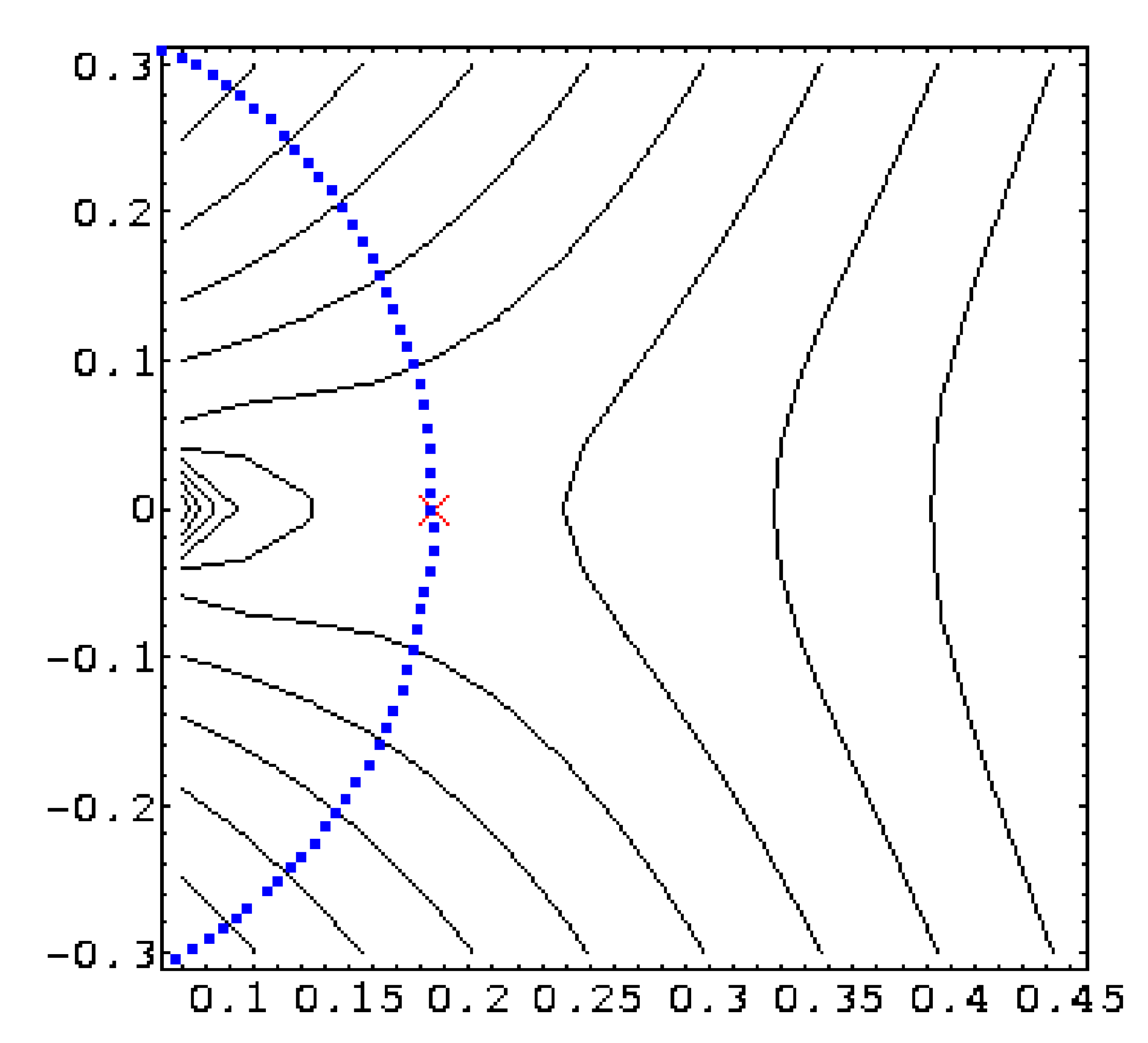}}
\hbox{
\vbox{\footnotefont\baselineskip6pt\narrower\noindent
Figure 6: Contour plot in the complex $N$ plane for the Mellin
inversion integral of $G(N,t)/G(N,t_0)$. The cross indicates 
the approximate location
of the real saddle, and the dotted line is the approximate
steepest descent path.
}}\hskip1truecm}
\endinsert
The effect of this $t$-factorisation is apparent comparing the contour
plot of fig.~1 with fig.~6, 
where we display 
$\Re\left[\xi N+\ln \left(G(N,t)/G(N,t_0)\right)\right]$ (all the
other parameters being as in fig.~1). 
In this case there is a saddle point on the real axis, 
while the complex saddle points 
have disappeared. The steepest descent contour goes over the real 
saddle, where the integrand is real, and there are no oscillations. 
In other words, whenever oscillations are present 
they are factorized in the initial condition, while the 
evolution factor $\Gamma(t,t_0;N)$~\aitwof\ has an inverse Mellin
transform which is 
smooth and asymptotically monotonic in $\xi$ as $\xi\to\infty$.

In conclusion, in order to
understand the impact of the running coupling resummation,
we give some estimates of the asymptotic behaviour 
of the inverse $N$--Mellin transform of
the evolution factor  $\Gamma(t,t_0;N)$, i.e. of the solution
$G(N,t)$ eq.~\aitwof\ when the boundary condition $\tilde
F_0(N,t)$ is constant.  If
$c>0$, the asymptotic behaviour of the evolution factor~\evfact\
is dominated by its rightmost singularity, which is  the 
simple pole  at $N=N_0(t_0)$,
where the Airy function in the denominator vanishes. 
The asymptotic behaviour is then the same as that of the splitting
function eq.~\splitas:
\eqn\gasy{G(\xi,t)\sim \as(t_0) 
x^{-N_0(t_0)}.}
Note that because $N_0(t)<\as(t) c$ the behaviour of the resummed evolution
factor is significantly softer than that which is found using the naive dual
anomalous dimension eq.~\aifive, namely $G\sim x^{-\as c}$ (see fig.~2). 
The same result is of course found by dominating the integral with the saddle
point displayed in fig.~6. 

If $c=0$ the rightmost singularity of the evolution factor is a branch
cut at $N=0$. In this case, the asymptotic behaviour is dominated by a
saddle point, located on the positive real axis at
\eqn\czsad{N_s= \left({k (t-t_0)\over\xi}\right)^{3/2},}
with $k= \Gamma(2/3)/
\Gamma(1/3) 3^{-2/3} (2\beta_0/\kappa)^{1/3}$.
This leads to 
\eqn\gasyz{G(\xi,t)\toinf{\xi} {1\over\sqrt{\xi^{5/2} \as(t)^{3/2}}}
\exp\left[  {k (t-t_0)^{3/2}\over \xi^{1/2}} \right].}
Because the leading asymptotics is constant, in this case we have also
given the first subleading correction, which corresponds to a 
power drop in $\xi$. This drop is somewhat milder than the $\xi^{-3/2}$
drop which is found~\sxap\ using the  naive dual
anomalous dimension eq.~\aifive.

Finally, if $c<0$ the behaviour can be estimated using a procedure
analogous to that which led to the corresponding
behaviour of the splitting function, eq.~\splitasneg, and in fact
it turns out to be the same:
\eqn\gasneg{G(\xi,t)\toinf{\xi}
\exp\left[ - \frac{4}{\sqrt{3}}
\left(\frac{-2c^3}{\kappa}\right)^{1/4}
\left({\xi\over\beta_0}\right)^{1/2}
\right].}
Notice that this asymptotic behaviour
is subdominant in comparison to that induced by a perturbative $N=0$
pole, such as is present in the standard one loop anomalous dimension. In
fact, if the boundary condition to perturbative evolution has a $N=0$
pole (corresponding to a constant behaviour in  $x$), 
the asymptotic behaviour of the evolution factor eq.~\gasy\ for
$N_0<0$ will be subdominant to that of the boundary condition.
It is, however, a slower drop in comparison to the power drop $G\sim
x^{-\as(t) c}$ which
would be found using the 
naive dual
anomalous dimension eq.~\aifive

\newsec{Asymptotic Behaviour and Resummation: the Generic Case}

The solution of the running coupling $\xi$--evolution equation with a
quadratic kernel studied in the previous section gives
the small $x$ asymptotic behaviour for generic
evolution kernels $\chi$ which have a minimum in the physical region
$0\le M\le1$. Consequently, it is possible to 
use this solution to resum
the perturbative running--coupling singularities.

The key observation which allows one to exploit the results of the
Airy resummation is  that the asymptotic small--$x$
behaviour of the solution is dominated by the form of the kernel in
the neighbourhood of its minimum, so we can view the quadratic kernel
eq.~\aione\ as the low-order truncation of the Taylor
expansion of the full
kernel about its minimum. Indeed, it was proven in ref.~\sxap\ that
if the running is included to finite perturbative order (or not at
all), then the coefficient of the leading behaviour in the splitting function at
each order is entirely determined by the value of the kernel and its
second derivative at the minimum, while higher--order derivatives
only affects
 terms which are asymptotically suppressed by inverse powers of
$\xi$. 

It is easy to check  that this remains true after resummation of
the running coupling effects (at least if $c\ge0$). To see this, assume
that the kernel eq.~\aione\ is corrected by a cubic term:
\eqn\phicub{\varphi(M)=c + {1\over 2} \chipp (M-\half)^2 + {1\over 6} \epsilon
(M-\half)^3.}
The solution eq.~\xthree\ can then be written as
\eqn\cubsol{G(N,t)= e^{- {1\over 6} \epsilon \frac{d^3}{dt^3}}
G^{(2)}(N,t),}
where $G^{(2)}(N,t)$ is the solution which corresponds to a quadratic
kernel $\varphi(M)$, given by eq.~\airysoln. The fact that the cubic
term only leads to a subleading correction to the asymptotic behaviour
is then a direct consequence of the fact that
\eqn\derzt{\frac{d}{dt}=\left(\frac{2\beta_0
N}{\kappa}\right)^{1/3}\frac{d}{dz}.}
Indeed, assume that the cubic term is treated
perturbatively. The change of asymptotic behaviour induced by this
term is then due to the fact that the position $z_0$ of the rightmost
zero of the Airy solution is shifted by the correction. The shifted
zero is located at $z_0+\delta$, where, expanding eq.~\derzt,
$\delta$ to first order is given by
\eqn\perzer{\delta ={1\over 6} \epsilon \frac{2\beta_0
N}{\kappa} {\Ai^{'''}(z_0)\over\Ai'(z_0)}.}

Hence, $\delta$ is proportional to $N$ and thus suppressed by inverse
powers of $\xi$ in the $x\to0$ limit, and so the leading behaviour
of the splitting function and solution 
as $x\to0$ are still asymptotically given by $x^{-N_0(t)}$, as in the
quadratic case
eqs.~\airysplit,\gasy. 
A similar line of argument leads to the general conclusion that the
leading asymptotic behaviour is not affected by corrections to the
quadratic approximation, although of course subleading corrections to
it are affected, as long as $c\ge0$. 
If $c<0$ these arguments fail, because the asymptotic behaviour is
dominated by the neighbourhood of the point $N\sim0$ (see
eq.~\splitasneg) i.e. (by duality) the neighbourhood of $M_0$ such
$\chi(M)=0$, and not by the minimum of $\chi$. However
in this case running coupling effects are subdominant in comparison to
the low--order perturbative terms and the boundary condition,
(compare eq.~\gasneg) and thus of no concern.

We can now exploit the universality of the quadratic result to resum
running coupling singularities, by matching the Airy anomalous
dimension to the singular anomalous dimension obtained at NLLx.
The first step in the procedure is to derive an
expansion of the Airy anomalous dimension~\aifive\ of the form
eq.~\gamexpone, namely, an expansion in powers of $\as(t)$ at fixed
$\as(t)/N$. Because
\eqn\zvsan{[z(\as,N)]^{-3/2}=
{1\over\sqrt{{2\over\kappa}\left(
{N\over\as}-c\right)}} \beta_0 \as {1\over\left(1-{\as c\over
N}\right)},}
the
large $z$ asymptotic 
formula in  eq.~\airyasymp\ can be used  to this purpose. 
The leading order term, already given in eq.~\aifive, is now
recognized as the leading order of the duality
expansion~\gamexpone\ of the Airy anomalous dimension:
\eqn\ailsexp{
\gamma_A(\as,N) = \frac{1}{2} - \sqrt{\frac{2}{\chipp}
\Big[\frac{N}{\as}-c\Big]}-\frac{1}{4} 
\frac{\beta_0\as}{1-\frac{\as}{N}c} + O(\as^2).
}  
So  the leading order term has the form 
that corresponds to $\varphi$ in eq.~\aione\ by the duality
relation eq.~\dual.  The term of order $\as\beta_0$ in eq.~\ailsexp\
is  the
lowest order effect of running: indeed,  it is equal
to $\Delta \gamma_{ss}$ eq.~\deltag,\deltagam\ for  the quadratic
$\varphi$ of eq.~\aione. In fact, 
the asymptotic expansion eq.~\airyasymp\ coincides with the asymptotic
expansion in powers 
of $N\beta_0/{\varphi^\prime}^3$ eq.~\sadexp,
obtained by evaluating by saddle--point the $M$--Mellin integral, 
so the expansion of the Airy function eq.~\gamexpone\ is recognized
to be the same as the running--coupling duality expansion of the
anomalous dimension eq.~\gamexpone, for the particular case of a
quadratic $\chi$ kernel. Hence, the perturbative effective kernel 
eq.~\chiefdef\ is an expansion of the Airy effective kernel
eq.~\efker. The expansion is factorially divergent but Borel summable.

Furthermore, the universality of the quadratic asymptotic behaviour
implies that the coefficient of the leading running--coupling
singularity to each order is entirely determined by knowledge of the
value of $\chi$ and its second derivative at the minimum. 
This can be seen explicitly at NLLx by considering the
running--coupling correction to the dual kernel $\Delta\chi_1$
eq.~\deltachi: whereas of course the function $\Delta\chi_1(M)$ for
generic $M$ is determined by the full $\varphi$, the residue of the
pole at $M=\half$ is entirely determined by knowledge of $c$, 
viewing  eq.~\aione\ as the quadratic truncation of the
Taylor series of $\chi$ about $M=\half$:
\eqn\laudel{\Delta\chi_1= {\beta_0\over 2}{c\over M-\half}+ O(M-\half).}
Therefore, the Airy effective kernel eq.~\efker\ is in fact a
(Borel) resummation of the leading running coupling corections to the kernel
eq.~\chiefdef\ for generic kernels with a minimum, and not only in the
particular case of a quadratic kernel.

It follows that we may resum the running coupling singularities
$\gamma_{\rm rc}$ eq.~\gamexpone\ by adding to the anomalous dimension
$\gamma$  the Airy 
anomalous dimension which corresponds to the quadratic approximation
to the kernel $\chi$ which is dual to $\gamma$, and
subtracting off the NLLx expansion eq.~\aifive\ in order to 
avoid double counting.
To NLLx it is in fact sufficient to determine the Airy resummation
from the $\chi$--kernel which is dual to the leading anomalous
dimension $\gamma_s$: explicitly, to this order
\eqn\resummedg{\eqalign{
\gamma_{\rm res}(\as,N) &= \gamma_s(\smallfrac{\as}{N})
+\as\gamma_{ss}(\smallfrac{\as}{N})
+\as\Delta\gamma_{ss}(\smallfrac{\as}{N})\cr 
&\qquad +\gamma_A(\as,N)-\half
+\sqrt{\smallfrac{2}{\chipp\as}[{N}-\as c]}+\quarter
\smallfrac{\beta_0\as N}{N-\as c},
}}
where $c$ is the value at its minimum of the $\chi$--function 
dual to $\gamma_s$.
Using eq.~\laudel\ and duality, the last term is seen to cancel 
the singularity in 
$\Delta\gamma_{ss}(\smallfrac{\as}{N})$, as it must since the 
singularities have been resummed in the Airy anomalous dimension
$\gamma_A$ eq.~\aithree.  

The resummed anomalous dimension $\gamma_{\rm res}$ eq.~\resummedg\ coincides
with the unresummed one up to terms which are N$^2$LLx, so the
resummation does not change the known low--order behaviour of the
anomalous dimension. In fact,  $\gamma_A$ is
rather small in comparison to the standard two-loop anomalous
dimension  (compare
fig.~4), thus it
only gives a rather small contribution 
to $\gamma_{\rm res}(\as,N)$ eq.~\resummedg: 
the main effect of the resummation procedure is the subtraction
of the running coupling singularities.
At asymptotically small $x$, the leading
behaviour of the unresummed $\gamma_s$ and $\gamma_{ss}$,
$x^{-\as c}$,  is replaced by the resummed behaviour ($x^{-N_0}$, if
$c>0$). However the subleading unresummed behaviour is not affected
by the resummation procedure. The subleading behaviour, due to terms
in the BFKL kernel beyond the quadratic approximation, is suppressed
by powers of $\xi$ in comparison to the leading one, i.e.  it is of the form
${1\over \xi^k} x^{-\as c}$. If these terms were
resummed, the subleading ${1\over \xi^k} x^{-\as c}$ behaviour would
be also replaced by a resummed one, $x^{-N_0+\delta(\xi)}$, with
$\delta\sim \xi^{- n}$, as in eq.~\perzer. This further subleading
running-coupling resummation is lacking in eq.~\resummedg, 
and it will become necessary when $x$ is so small
that the shape of the anomalous dimension around $N\sim \as c$ is important.
\topinsert
\vbox{
\epsfxsize=9.8truecm
\centerline{\epsfbox{reschi.ps}}
\hbox{
\vbox{\footnotefont\baselineskip6pt\narrower\noindent
Figure 7: The leading-- (dotted) 
and next--to--leading order (solid)  double--leading $\chi$ kernels compared
to the resummed effective kernel (dashed). All curves are computed with
$\as=0.2$ in the R--resummation
scheme of ref.~\sxphen.
}}\hskip1truecm}
\endinsert

By using the resummed anomalous dimension
eq.~\resummedg\ in the duality relation eq.~\dual, we 
may finally determine a resummed effective $\chi_{\rm
eff}$ eq.~\chiefdef. The effect of this resummation is shown in
fig.~7, where we compare the effective $\chi$ determined at leading
and next--to--leading order of the double--leading
expansion~\refs{\sxres,\sxphen}, and the resummed next--to--leading
result. The NLO kernel is affected by the singularity, 
which is then removed by the
resummation. Note that the resummed result shown in this figure is
defined as the dual~\dual\ of the resummed $\gamma$, obtained by
combining the Airy resummation eq.~\resummedg\ with the
double--leading expansion of the anomalous which we discussed in
refs.~\refs{\sxres,\sxphen} (see fig.~8 below). For ease of
comparison with our previous work, the double--leading kernel shown
here is the same as that displayed in fig.~2 of ref.~\sxres, which
violates the   momentum conservation constraint  $\chi(0)=1$ by
sub-subleading terms. The
resummation procedure is then applied to this kernel and thus leads to a
result which also does not exactly conserve momentum.

In \MS--like schemes the running coupling singularities are actually
shifted into the quark sector, so the resummation should instead be 
performed in the quark sector. This can be done by noting that the
running coupling singularities can be removed from the anomalous
dimension $\gamma^+$
by redefining the normalization of the gluon distribution by a factor
$u[\gamma_s(\as/N)]$. Such a redefinition amounts to a conventional
change of factorization scheme provided only that
$u[\gamma(0)]=1$~\refs{\mom,\sxphen}.  Upon this redefinition 
the large eigenvalue
of the anomalous dimension matrix $\gamma^+$ (which so far we
identified with the anomalous dimension, recall eq.~\Gdef) and the
quark sector anomalous dimension $\gamma_{qg}$
change according to~\refs{\mom,\sxphen}
\eqn\schch{\eqalign{&\gamma^+\to
{{\gamma}^+}^\prime=\gamma^++\beta_0{\chi_0(\gamma_s)\over\chi_0'(\gamma_s)}
{d\over dM} \left[\ln u(M)\right]\bigg|_{M=\gamma_s}\cr
&\gamma^{qg}\to{\gamma^{qg}}^\prime=u^{-1}(\gamma_s)\gamma^{qg}.\cr}}
The choice
$u(\gamma_s)={\sqrt{-\chi_0'(\gamma_s)}\over N/\as}$ 
is then sufficient to remove the
singular contribution eq.~\deltag\ from $\gamma^+$ (while the factor
in the denominator ensures that this is a {\it bona fide} scheme
change). The scheme change, however, introduces a
square-root singularity in the quark anomalous dimension
$\gamma^{qg}$:  indeed, we get \eqnn\schchg\eqnn\schchq
$$\eqalignno{&
{\gamma^+}^\prime=\gamma^++
\beta_0\left(
\frac{\chi_0''(\gamma_{s})\chi_0(\gamma_{s})}
{2\chi_0'^2(\gamma_{s})}-1\right)&\schchg\cr
&{\gamma^{qg}}^\prime={\chi_0(\gamma_s)\over\sqrt{-\chi_0^\prime
(\gamma_s)}}\gamma^{qg}.&\schchq\cr}
$$
 The transformation which takes one to the \MS\ scheme~\CH\ 
is instead  given by a factor $u(M)=R(M)=
r(M)/\sqrt{\chi_0^\prime}$, where $r(M)$ is a regular function of $M$.
Now, we can perform the desired running coupling resummation in the
quark sector by starting with the
resummed anomalous dimension eq.~\resummedg, and then performing a
scheme change of the form eq.~\schch, while demanding  
that ${\gamma^+}^\prime$ be still given by eq.~\schchg. The
resummation is thus shifted entirely into the quark sector, and 
${\gamma^{qg}}^\prime$ is no longer given by eq.~\schchq, but rather
by
\eqn\resummedqg{ {\gamma^{qg}}^\prime(\as,N)= \Ai[z(\as(t),N)]e^{{2\over3}
\left[z(\as,N)\right]^{-3/2}} \gamma^{qg}(\as,N),}
which shows explicitly that the square--root singularity in
${\gamma^{qg}}^\prime$ eq.~\schchq\ 
is eliminated by the resummation (the extra exponential
factor removes the term corresponding to $\gamma_s$ in
the expansion of the Airy resummed contribution).

This running--coupling
resummation can finally be combined with the small--$x$ resummation of
the anomalous dimension previously discussed by
us~\refs{\sxres,\sxphen}. In such case, the expansion of the anomalous
dimension in powers of $\as$ at fixed $\as/N$ is obtained by duality
from a reorganized expansion of the $\chi$ kernel, where in particular
the leading--order term is 
\eqn\tilchidef{\as\tilde\chi_0=\as\chi_0+\Delta\lambda,} 
with
$\Delta \lambda=\as^2\chi_1(1/2)+\dots$  such that 
$\lambda\equiv\as\chi_0(1/2)+\Delta\lambda$
coincides with the value  of the all-order
$\chi$ at its minimum. Consequently,  
$\lambda$ must be treated as a free parameter.
To resum the corresponding NLLx running coupling singularity it is
sufficient to identify the quadratic $\varphi$ eq.~\aione\ with the
expansion of $\tilde\chi_0$ about its minimum, so that $c=\lambda/\as$,
while $\kappa$ is given by eq.~\kapval, and then resum according to 
eq.~\resummedg. This resummed NLLx anomalous dimension can then be
combined with the standard two--loop anomalous
dimension~\refs{\sxres,\sxphen} to give a
fully resummed double--leading expansion.

The phenomenological impact of these manipulations can be understood
on the basis of the asymptotic estimates discussed in the previous
section. 
If $c>0$ the running coupling resummation shifts the rightmost
singularity of the small--$x$ anomalous dimension 
from  $N=\as(t)c$ to the smaller value $N=N_0(t)$:
$N_0(t)<\as(t)c$, as discussed in the previous section (see
eq.~\aitwob). The
asymptotic behaviour is thus somewhat softened. 
If $c=0$ the running coupling resummation has a modest
effect, in that it simply turns the square--root branch point of the
unresummed naive dual anomalous dimension eq.~\aifive\ into a cubic
branch point; in fact, when $c=0$ there are no running coupling
singularities, since in this case $\Delta\chi_1$~\deltachi\ and its
higher--order generalizations are regular as $M\to\half$~\sxphen.
Finally, if $c<0$, both the unresummed and resummed 
behaviours
are 
subdominant in comparison to the perturbative behaviour, and thus the
phenomenological impact of the resummation is negligible. 

\topinsert
\vbox{
\epsfxsize=9.8truecm
\centerline{\epsfbox{resgam.ps}}
\hskip4truecm\hbox{
\vbox{\footnotefont\baselineskip6pt\narrower\noindent
Figure 8: The double--leading anomalous dimensions with $c=1$ and
$\as=0.2$ 
(i.e. $\lambda=0.2$~\sxphen) computed to NLO in the R--resummation
scheme of ref.~\sxphen\  with \MS\ factorization (dotted). The same
after the scheme change eq.~\schchg\ (solid), displaying the
running--coupling singularity, and then after the singularity is
removed by resummation~\resummedg\ (dashed). 
}}\hskip1truecm}
\endinsert
This situation is displayed in fig.~8, with the phenomenologically
plausible~\sxphen\ value $c=1$. In this figure, 
we show the NLO double--leading
anomalous dimension $\gamma^+$ in the \MS\ scheme (used in
ref.~\sxres), and we compare it to  the asymptotically unstable
anomalous dimension obtained from it performing in reverse the scheme change
eq.~\schchg, so the running coupling singularity is moved  back into
$\gamma^+$. 
Note that 
the momentum
conservation constraint $\gamma^+(1)=0$ is exactly enforced on all
the anomalous dimensions shown in fig.~8, by means of a sub-subleading
subtraction, as discussed in ref.~\sxres.
The scheme change has a very small effect for almost all
values of $N$, but it generates a strong instability in a small region
of $N$, $0.2<N\lsim 0.23$. (recall that the anomalous dimension has a
branch cut at $N=\lambda$) .
Finally, we show the result of applying the resummation
eq.~\resummedg\ to this unstable anomalous dimension:
the instability is completely  removed, and the anomalous dimension
remains smooth for all $N>\lambda$. The
 softening of the small $N$
behaviour in the vicinity of the branch cut appears to be marginal, as 
expected because of the
smallness of the $\gamma_A$ contribution to the resummed anomalous
dimension eq.~\resummedg (see fig.~4). 
Hence, the main
effect  of the resummation is to remove the instability, thereby
widening the region of validity of the resummed anomalous dimension
down to $N=\lambda$. Below this value, a further subleading
running--coupling resummation would be required in order to recover the
correct asymptotic $x^{-N_0}$ behaviour. 
The phenomenological effects of the instability and its resummation
turn out to be
small in the HERA region because of the limited range of $N$ in which
it appears, but they may become relevant at
yet higher energies~\sxphen. 

The running coupling resummation can be generalized to higher orders
by using the expansion eq.~\rcgexp\ in the running--coupling evolution
equation, to generate recursive equations of the form of eq.~\nloeq,
which determine each $G_i$ in terms of the lower--order terms.
Note that eq.~\nloeq\ is obtained by neglecting higher order
corrections which correspond to commutators of $\ash$  and
$\chi_0$. This terms are nonsingular in the asymptotic limit, but will
have to be included explicitly into the kernel when going to yet
higher orders.
It is clear from eq.~\nloeq\ that these subleading corrections remain
perturbative, in the sense that no further singular contributions to
$\chi$  are generated
beyond those which appear at the leading level. However, the
coefficients of these singularities will receive subleading corrections
which can be determined by explicit solution of eq.~\nloeq\ and its
higher order generalizations.

\newsec{Conclusions}

In this paper, we 
have shown how to introduce the running of the coupling to all orders in a
BFKL--like $x$--evolution equation, in a way which is consistent with
standard renormalization--group evolution. 
We have thus been able to establish three main results.

First, we  have
proven that the solution to the running coupling BFKL equation 
also satisfies
a renormalization--group equation, with factorized boundary conditions
and a scale dependence entirely determined by an anomalous dimension
which is related by duality to the BFKL kernel.
This generalizes to the case of all--order running coupling the
duality previously derived by us in perturbation theory.
Second, by using the well--known Airy solution of  this 
running--coupling BFKL equation with a
quadratic kernel, we have shown that the 
 resummed splitting function is free of the 
unphysical oscillations which characterize the small $x$
behaviour of the solutions to
the running--coupling BFKL equation, since these get
factorized into the initial condition.
Finally, we have shown how the Airy solution can be used  to 
construct a general resummation of running--coupling singular terms,
thanks to the fact that the 
quadratic kernel provides
an approximation to the full kernel which correctly determines the
asymptotic small $x$ behaviour. 
The net effect of the resummation is small in phenomenologically
interesting regions, but it removes a perturbative instability which
spoiled the asymptotic small $x$ behaviour of the unresummed results.

In summary, even
though it may turn out that in the Regge limit   the perturbative
behaviour is overwhelmed by
power--suppressed contributions, or corrections which
go beyond perturbation theory, 
we have shown that this need not necessarily be so: 
a leading--twist
perturbative approach may provide a consistent
description of scaling violations in the asymptotic
small $x$ limit.
\medskip
{\bf Acknowledgements}: We thank M.~Ciafaloni and
                        G.~Salam for various discussions.
This work was supported in part by
EU TMR  contract FMRX-CT98-0194 (DG 12 - MIHT).
\bigskip
\footatend\vfill\supereject\immediate\closeout\rfile\writestoppt
\baselineskip=14pt\centerline{{\bf References}}\bigskip{\frenchspacing%
\parindent=20pt\escapechar=` \input refs.tmp\vfill\eject}\nonfrenchspacing
\vfill\eject
\bye